\documentclass[times,sort&compress,3p]{elsarticle} 
\journal{Journal of Multivariate Analysis}
\usepackage{natbib}
\usepackage{lineno,hyperref}
\usepackage{amsmath,amssymb,enumerate,amsthm}
\usepackage{latexsym}
\usepackage{color,xcolor}
\usepackage{graphicx,color}
\usepackage{adjustbox}
\usepackage{amsfonts}
\usepackage{caption}
\usepackage{setspace}
\usepackage{txfonts}
\usepackage{subcaption}
\usepackage{tabularx}

\newtheorem{theorem}{Theorem}

\newtheorem{algorithm}{Algorithm}

\newtheorem{remark}{Remark}

\def\BB{\mathbf B}
\def\LL{\mathbf L}
\def\XX{\mathbf X}

\begin{document}
\begin{frontmatter}
	
\title{Estimating tail probabilities of the ratio of \\
the largest eigenvalue to the trace of a Wishart matrix}
\author[rvt]{Yinqiu He}
\ead{yqhe@umich.edu}
\author[rvt]{Gongjun Xu\corref{cor1}}
\ead{gongjun@umich.edu}

\cortext[cor1]{Corresponding author}
\address[rvt]{Department of Statistics, University of Michigan, Ann Arbor, MI, USA}

\begin{abstract}
This paper develops an efficient Monte Carlo  method to estimate the tail probabilities of the ratio of the largest eigenvalue to the trace of the  Wishart matrix, which plays an important role in multivariate data analysis. 
The estimator is constructed based on a change-of-measure technique and  it is proved to be  asymptotically  efficient for both the real and complex Wishart matrices. 
Simulation studies further show the improved performance of the proposed method over  existing approaches based on asymptotic approximations, especially when estimating probabilities of rare events.


 \end{abstract}

\end{frontmatter}

\section{Introduction} \label{sec:intro}
Consider $n$ independent and identically distributed (iid) $p$-dimensional observations $\mathbf{x}_1 ,  \ldots, \mathbf{x}_n$ from 
 a real or complex valued Gaussian distribution with mean zero and covariance matrix $\Sigma =\sigma^2I_p$. 
  Here $\sigma^2$ is an unknown scaling factor and $I_p$ is the $p\times p$ identity matrix. Define the $n\times p$ data matrix $\mathbf{X}= (\mathbf{x}_1, \ldots, \mathbf{x}_n )^\top$, and assume $\lambda_{1}\geq \cdots\geq \lambda_{p}$ are the ordered 
 real eigenvalues of  the sample covariance matrix $\hat\Sigma=  \XX^H \XX/n$, where $^H$ denotes the conjugate transpose. Note that if $p>n$, the last $p-n$ of the $\lambda$s are zero.
Let  $U_{n,p}$ be the ratio of the largest eigenvalue to the trace, viz.
\begin{eqnarray} \label{def:u}
	U_{n,p}=\frac{\lambda_{1}}{(\lambda_1 + \cdots + \lambda_p)/\min (n,p)}.
\end{eqnarray}
We are interested in estimating the rare-event tail probability 
 $\alpha_{n,p}(x) = \Pr\left(U_{n,p} > x\right),$
where $x$ is some constant such that $\alpha_{n,p}(x)$ is small.
 Estimating rare-event tail probabilities is often of interest in multivariate data analysis. 
For instance, in multiple testing problems, 
 it is often needed to evaluate very small $p$-values for  individual test statistics to control the overall false-positive error rate. 
 
The random variable $U_{n,p}$ plays an important role in multivariate statistics when testing the covariance structure. For instance, it has been  used to test for equality of the population covariance to a scaled identity matrix, viz.
$$\mathcal{H}_0:\Sigma =\sigma^2 I_p \quad \mbox{vs.} \quad \mathcal{H}_1:\Sigma \neq\sigma^2 I_p
$$ 
with $\sigma^2$ unknown, i.e., the so-called sphericity test; see, e.g., \cite{muirhead2009aspects}.
The test statistic $U_{n,p}$ does not depend on the unknown variance parameter $\sigma^2$ and has high detection power against alternative covariance matrices with a low-rank perturbation of the null $\sigma^2I_p$. 
In particular, under the alternative of rank-1 perturbation with  $\Sigma=hh^\top +\sigma^2 I_p$ for some unknown $h\in\mathbb R^p$ and $\sigma^2$, 
the likelihood ratio test statistic $\mathcal{L}_n ={\sup_{h,\sigma^2}f_1(X; h,\sigma^2)}/{\sup_{\sigma^2}f_0(X; \sigma^2)}$
can be written as a monotone function of  $U_{n,p}$   and therefore  $\alpha_{n,p}(x)$ corresponds to the $p$-value; see, e.g., \citep{muirhead2009aspects,bianchi2011performance}.  
Please refer to \cite{krzanowski2000principles, muirhead2009aspects,paul2014random}  for more discussion and many other applications. 

The exact distribution of $U_{n,p}$ is difficult to compute, especially when estimating rare-event tail probabilities. 
Note that $\XX^H\XX/(n\sigma^2)$ follows a Wishart distribution $\mathcal{W}_{\beta,p}(n, I_p/n)$, with $\beta=1$ for real Gaussian and $\beta=2$ for complex Gaussian. So the distribution of  $U_{n,p}$ corresponds  {to that of the ratio of the largest eigenvalue to the trace} of a $\mathcal{W}_{\beta,p}(n, I_p/n)$. However, this distribution is nonstandard and  {exact formulas based on it typically involve} high-dimensional integrals or inverses of Laplace transforms. 
Numerical  evaluation has been studied in  \cite{davis1972ratios,schuurmann1973distributions,kuriki2001tail,kortun2012distribution,wei2012exact,chiani2014distribution}.
But for high-dimensional data with large $p$, the computation becomes more challenging, which is notably the case when $\alpha_{n,p}(x)$ is small, due to the additional  computational  cost to control the relative estimation error of $\alpha_{n,p}(x)$. 
  
The asymptotic distribution of $U_{n,p}$ with $p$ and $n$ both going to infinity has also been studied in the literature.  
It is known that  $U_{n,p}$ asymptotically behaves similarly to the largest eigenvalue $\lambda_1$, whose limiting distribution has been studied in \cite{johansson2000shape, johnstone2001},  and $U_{n,p}$ also asymptotically  follows the Tracy--Widom distribution; see, e.g., \citep{bianchi2011performance,nadler2011distribution}. That is, 
\begin{equation}\label{tw}
	\Pr\left(\frac{U_{n,p}-\mu_{n,p}}{\sigma_{n,p}}>x\right) \to 1-\mathcal{TW}_{\beta}(x),
\end{equation}
where $\mathcal{TW}_{\beta}$ denotes the Tracy--Widom distribution of order $\beta$, with $\beta=1$ or $2$ for real and complex valued observations, respectively. In particular, for real-valued observations, the centering and scaling constants 
\begin{eqnarray} \label{parameter}
\mu_{n,p} = \frac{1}{n}\left(\sqrt{n-\frac{1}{2}}+\sqrt{p-\frac{1}{2}}\right)^2 \mbox{ and }
\sigma_{n,p} = \frac{1}{n}\left (\sqrt{n-\frac{1}{2}}+\sqrt{p-\frac{1}{2}}\right)\left(\frac{1}{\sqrt{n-{1}/{2}}}+\frac{1}{\sqrt{p-{1}/{2}}}\right)^{1/3}
\end{eqnarray} lead to a convergence rate    of the order $O \{ \min (n,p)^{-2/3}\}$; see \citep{ma2012accuracy}.
  For the complex case, similar expressions can be found in \cite{karouicomplexrate}.  Nadler~\cite{nadler2011distribution}  studied  the accuracy of  the Tracy--Widom approximation for finite values of $n$ and $p$. He found that   the approximation  may be inaccurate for small and even moderate values of $p$ when $n$ is large. Therefore, he proposed a correction term to improve the approximation result, which is derived using the Fredholm determinant representation, and he showed that  the approximation rate is $o\{ \min (n,p)^{-2/3}\}$ when $X$ follows a {\it complex} Gaussian distribution. 
In the {\it real} Gaussian case, which is of interest in many statistical  applications, Nadler~\cite{nadler2011distribution}  conjectured that the result also holds.
The calculation of the correction term in \cite{nadler2011distribution} depends on the second derivative of the non-standard Tracy--Widom distribution, which usually involves a numerical discretization scheme.
 
Another limitation of the existing methods is that they may become less efficient when estimating   small tail probabilities of rare events.
This paper aims to address this rare-event estimation problem. 
In particular, we propose  an efficient Monte Carlo method to estimate the exact tail probability of $U_{n,p}$ by utilizing importance sampling. 
 {The latter} is a commonly used tool to reduce Monte Carlo variance and it has been found helpful to estimate small tail probabilities, especially when the event is rare, in a wide variety of stochastic systems with both light-tailed and heavy-tailed distributions; see, e.g.,
 \citep{SIE76,AsmKro06,DupLedWang07,ASMGLY07,BlaGly07,LiuXu12,LiuXuTomacs,xu2014rare}.
 
An importance sampling algorithm needs to construct an alternative sampling measure (a change of measure) under which the eigenvalues are sampled.
Note that it is necessary to normalize the estimator with a Radon--Nikodym derivative to ensure an unbiased estimate.
Ideally, one develops a sampling measure so that the event of interest is no longer rare under the sampling measure.
The challenge is of course the construction of an appropriate sampling measure, and one common heuristic is to utilize a sampling measure that approximates the conditional distribution of $U_{n,p}$ given the event $\{U_{n,p}>x\}$.
This paper proposes a change of measure $Q$ that asymptotically  approximates the conditional measure   $\Pr(\cdot \mid U_{n,p} >x)$. 
We carry out a rigorous  analysis  of the proposed estimator for $U_{n,p}$ and show that it is 
{asymptotically  efficient}.
Simulation studies  show that the proposed  method outperforms  existing approximation approaches, especially when estimating probabilities of rare events.

The remainder of the paper is organized as follows. In Section~\ref{description}, we propose our importance sampling estimator and establish its asymptotic efficiency in Theorem~\ref{thm:mainthm}.  Numerical results are presented in Section~\ref{sec:num} to illustrate its performance. We  discuss the possibility of generalizing the result to the ratio of the sum of the largest $k$ eigenvalues to the trace of a Wishart matrix in Section~\ref{sec:ext}. The proof  of Theorem~\ref{thm:mainthm} is given in Section~\ref{sec:proof}. 

\section{Importance sampling estimation} 
\label{description}

For ease of discussion, we  consider the setting $p\leq n$, $p\to\infty$ and $n\to\infty$. When $p>n$, the algorithm and theory are essentially the same up to switching labels of $p$ and $n$, which is explained in Remark \ref{rm:pnswitch}. We use the notation $\beta$ to denote  the real Wishart Matrix ($\,\beta=1$) and  complex Wishart matrix ($\,\beta=2$). 
Since $U_{n,p}={p\lambda_{1}}/(\lambda_1 + \cdots + \lambda_p)$  is invariant to $\sigma^2$, the analysis does not depend on the specific values of $\sigma^2$, and we take  $\sigma^2$ as follows in order to simplify the notation and unify the real and complex cases under the same representation, as specified in Eq.~\eqref{originaldensity} below: 
\begin{itemize}
	\item [a)]
	When $\beta=1$, we assume  that $\sigma^2=1$.  That is, the entries of $\mathbf{X}$  are iid  $\mathcal{N}(0,1)$, and $\lambda_{1},\ldots, \lambda_{p}$ are the ordered eigenvalues of ${\mathbf{X}^{\top} \mathbf{X}/n }$. 
	\item [b)]
	When $\beta=2$, we assume $\sigma^2=2$. We consider the circularly symmetric Gaussian random variable  \citep{tse2005fundamentals},
 	  and  we  write $ X =Y+iZ  \sim \mathcal{CN}(0,\sigma^2)$ when $Y$ and $Z$ are iid  $\mathcal{N}(0,{\sigma^2}/{2})$. 
 	  In the following, we assume that the entries of $\mathbf{X}$ are iid  $\mathcal{CN}(0,2)$, and that $\lambda_{1},\ldots, \lambda_{p}$ are the ordered eigenvalues of ${\mathbf{X}^{H} \mathbf{X}/n}$.

\end{itemize}
As mentioned, e.g., in \citep{dumitriu2002matrix}, the $p$  eigenvalues  $\lambda_{1}\geq \cdots\geq\lambda_{p}\geq 0$ are distributed
with probability density function
 \begin{eqnarray}\label{originaldensity}
	f_{n,p,\beta}(\lambda)= C_{n,p,\beta} \prod_{i<j}^{p} | \lambda_i-\lambda_j  |^{\beta} \prod_{i=1}^{p} \lambda_i^{{\beta (n-p+1)}/{2} -1} e^{-n(\lambda_1 + \cdots + \lambda_p)/2},
\end{eqnarray}
when $\beta \in \{ 1, 2\}$, where $C_{n,p,\beta}$ is a normalizing constant given by
\begin{equation*}\label{cnp}
C_{n,p,\beta}=p!\left(\frac{n}{2}\right)^{ {\beta np}/{2}}\prod_{j=1}^p \frac{\Gamma(1+{\beta}/{2})}
{\Gamma(1+ {\beta j }/{2})\Gamma \{ \beta (n-p+j)/2\}}.
\end{equation*}
 Then the target probability $\alpha_{n,p}(x)=\Pr \left(U_{n,p} > x\right)$ can be written as  
 $$
 \alpha_{n,p}(x) =\int_{\lambda_1\geq\cdots\geq\lambda_p\geq 0} \mathbf{1}{(U_{n,p} > x)} f_{n,p,\beta}(\lambda_{1},\ldots,\lambda_{p})d \lambda_1 \cdots d\lambda_p,
 $$
 where $\mathbf{1}$ is the indicator function. 
As discussed in the Introduction, direct evaluation of the above $p$-dimensional integral is computationally challenging, especially when $p$ is relatively large.

This work aims to design an efficient Monte Carlo method to estimate $\alpha_{n,p}(x)$. We first introduce some computational concepts from the rare-event analysis  literature, which helps to evaluate the computation efficiency of a Monte Carlo estimator.  

Consider an estimator $L_{n,p}(x)$ of a rare-event probability $\alpha_{n,p}(x)$, which goes to 0 as $n\to\infty$. We  simulate $N$ iid ~copies of $L_{n,p}(x)$, say $L_{n,p}^{(1)}(x), \ldots, L_{n,p}^{(N)}(x)$, and  obtain the average  estimator 
$\bar L_{n,p}(x) = \{ L_{n,p}^{(1)}(x) + \cdots + L_{n,p}^{(N)}(x)\}/N$.
We want to  control the relative error $|\bar L_{n,p}(x) - \alpha_{n,p}(x)|/\alpha_{n,p}(x)$ such that for some prescribed $\varepsilon, \delta \in (0, \infty)$,
\begin{equation*}
\Pr   \{ |\bar L_{n,p}(x) - \alpha_{n,p}(x)|/\alpha_{n,p}(x)> \varepsilon \}< \delta.
\end{equation*}
Consider the direct Monte Carlo estimator as an example. 
The  direct Monte Carlo directly generates samples from the density \eqref{originaldensity} and uses $L_{n,p}(x) = \mathbf{1}(U_{n,p}>x).$ So in each simulation we have a Bernoulli variable with mean $\alpha_{n,p}(x)$. According to the Central Limit theorem,  the direct Monte Carlo simulation requires
$N=\Theta\{ \varepsilon^{-2}\delta ^{-1} \alpha_{n,p}(x)^{-1}\}$ iid  replicates to achieve the above accuracy, where the notation $\Theta$ is defined as follows. For any $a_n$ and $b_n$ depending on $n$,  $a_n=\Theta(b_n)$ means that $0 < \liminf_{n\to\infty}|a_n/b_n|\leq \limsup_{n\to\infty}|a_n/b_n|<\infty.$
This implies that the direct Monte Carlo method becomes inefficient and even infeasible as $\alpha_{n,p}(x)\to 0$.

A more efficient estimator is the {asymptotically efficient} estimator; see, e.g., \citep{SIE76,AsmKro06}. An unbiased estimator $L_{n,p}(x)$ of   $\alpha_{n,p}(x)$ is called  asymptotically efficient if 
\begin{equation}\label{logeff}
\liminf_{n\to\infty} { \ln [\mathrm{var}\{L_{n,p}(x)\}] }/{  \ln \{ \alpha_{n,p}(x)^2 \}} \geq  1.
\end{equation}
Note that \eqref{logeff} is equivalent to 
\begin{equation}\label{logeff1}
\limsup_{n\to\infty} { \mathrm{var}\{ L_{n,p}(x) \} }/{\alpha_{n,p}(x)^{2-\eta}} =0, \end{equation}
for any $\eta >0$.  In addition, since $\mathrm{E}(L_{n,p}^2)\geq {\rm var}\{L_{n,p}(x)\}$ and 
$$
\limsup_{n\to\infty} {\ln \{ \mathrm{E}(L_{n,p}^2) \} }/
{\ln \{\alpha_{n,p}(x)^2 \} }\leq 1
$$ 
by H\"{o}lder's inequality, 
\eqref{logeff} is also equivalent to 
\begin{equation*}\label{eqv2}\lim_{n\to\infty}  {\ln \{\mathrm{E}(L_{n,p}^2 )\}}/{\ln \{\alpha_{n,p}(x)^2 \} }=1.
\end{equation*}  

When $L_{n,p}(x)$ is  asymptotically efficient,  by Chebyshev’s inequality, 
$$
\Pr   \{ |\bar L_{n,p}(x) - \alpha_{n,p}(x)|/\alpha_{n,p}(x)> \varepsilon \} \leq \mathrm{var} \{ L_{n,p}(x)\} / \{ N\alpha_{n,p}(x)^2\epsilon^2\},
$$
 and therefore  \eqref{logeff1}  implies that we only need  $N=O\{ \varepsilon^{-2}\delta ^{-1}\alpha_{n,p}(x)^{-\eta} \}$, for any $\eta>0$, iid  replicates of $L_{n,p}(x)$.
Compared with the direct Monte Carlo simulation, 
 efficient estimation substantially reduces the computational cost, especially when $\alpha_{n,p}(x)$ is small.
  
To construct an asymptotically efficient estimator, we use the importance sampling technique, which is an often used method for variance reduction of a Monte Carlo estimator. We use $P$ to denote the probability measure  of the eigenvalues $\lambda_1,\ldots,\lambda_p$.
The importance sampling estimator is constructed  based on the identity
$$
\Pr (U_{n,p} > x) = \mathrm{E} \{\mathbf{1}{(U_{n,p} > x)}\}
=\mathrm{E}_Q \left\{\mathbf{1}{(U_{n,p} > x)}\,{dP}/{dQ}\right\},
$$
where 
$Q$ is a probability measure such that the Radon--Nikodym derivative $dP/dQ$ is well defined on the set $\{U_{n,p} > x\}$, and we use  $\mathrm{E}$ and $\mathrm{E}_Q$ to denote the expectations under the measures $P$ and $Q$, respectively.
Let $f^Q_{n,p}$ be the density function of  the eigenvalues $\lambda_1,\ldots,\lambda_p$ under the change of measure $Q$.
Then, the random variable defined by
\begin{equation*}\label{estr}
L_{n,p}= \mathbf{1}{(U_{n,p} > x)}  {f_{n,p}(\lambda_1,\ldots,\lambda_p) }/{f^Q_{n,p}(\lambda_1,\ldots,\lambda_p)}  
\end{equation*}
is an unbiased estimator of $\alpha_{n,p}(x)$ under the measure $Q$. 
Therefore, to have $L_{n,p}$  asymptotically efficient, we only need to choose a change of measure $Q$ such that  
\begin{equation}\label{logeff2}
 \liminf_{n\to\infty}\frac{1}{ |2\ln\alpha_{n,p}(x)|} \,  |\ln \mathrm{E}_Q \{ \mathbf{1}{(U_{n,p} > x)}  {f_{n,p}(\lambda_1,\ldots,\lambda_p)^2}/{f^Q_{n,p}(\lambda_1,\ldots,\lambda_p)^2}  \}\ |\geq  1.
\end{equation}

To gain insight into the requirement \eqref{logeff2}, we consider some examples. 
First consider the direct Monte Carlo with $f^Q_{n,p} =  f_{n,p}$; the right-hand side of \eqref{logeff2} then equals $1/2$ which is smaller than 1. 
On the other hand, consider
$Q $ to be  the conditional probability measure given $U_{n,p} > x$, i.e., $f^Q_{n,p}(\cdot) = f_{n,p}(\cdot ) \mathbf{1}(U_{n,p} > x)/\alpha_{n,p}(x)$; then  the right-hand side of \eqref{logeff2} is exactly 1.
Note that this change of measure is of no practical use since $L_{n,p}$ depends on the unknown $\alpha_{n,p}(x)$. 
But  if we can find a measure $Q$ that is a good approximation of the conditional probability measure given $U_{n,p} > x$, we would expect \eqref{logeff2} to hold and the corresponding estimator $L_{n,p}$   to be efficient.
In other words, the asymptotic efficiency criterion requires the change of measure $Q$ to be a good approximation of the conditional distribution of interest. 

Following the above argument, we construct the change of measure $Q$ as follows, which is motivated by a recent study of Jiang et al. ~\cite*{xu2016rare}. These authors studied the tail probability of the largest eigenvalue, i.e., $\Pr(\lambda_1>px)$ with $p>n$ and  proposed a change of measure that approximates the conditional probability measure given $\lambda_1>px$ in total variation when $p\gg n$. 
It is known  that  the asymptotic behaviors of $\lambda_1$ and $U_{n,p}$ are closely related.  
We therefore adapt the change of measure to the current problem of estimating $U_{n,p}$. 
However, we would like to clarify that the problem of estimating $U_{n,p}$ is different from that in \cite{xu2016rare} in terms of both  theoretical justification and  computational implementation,  which is further discussed in Remark \ref{rm:compare}. 
 
Specifically, we propose the following importance sampling estimator.
\begin{algorithm}\label{algo:1}
\em
Every iteration in the algorithm contains three steps, as follows:
\begin{itemize}\label{algorithm1}
\item[Step\,1] We use the matrix representation of the $\beta$-Laguerre ensemble introduced in \cite{dumitriu2002matrix}, and generate the matrix
$\LL_{n-1,p-1,\beta} =\BB_{n-1,p-1,\beta} \BB_{n-1,p-1,\beta}^\top,$
where $\BB_{n-1,p-1,\beta}$ is a bidiagonal matrix defined by
\begin{eqnarray*}
\BB_{n-1,p-1,\beta}=\left(\begin{array}{ccccc}
\chi_{ \beta(n-1)} &\\
\chi_{\beta(p-2)}& \chi_{\beta(n-2)}&\\~\\
 &\ddots&\ddots\\~\\
&&\chi_{\beta}& \chi_{\beta \{ n-(p-1)\} }&\\
\end{array}\right)_{(p-1)\times(p-1)}.
\end{eqnarray*}
The notation $\chi_a$ denotes the square root of the chi-square distribution with $a$ degrees of freedom, and
the diagonal and sub-diagonal elements of $\BB_{n-1,p-1,\beta}$ are generated independently. We then compute the corresponding ordered eigenvalues  of $\LL_{n-1,p-1,\beta}/n$, denoted by $\lambda_{2}\geq \cdots\geq \lambda_{p}$.
\item[Step\,2] Conditional on $\lambda_{2},\ldots,\lambda_{p}$, we sample $\lambda_{1}$ from an exponential distribution with density
\begin{eqnarray} \label{eq:expodist}
f(\lambda_{1})=  nr e^{-nr \left(\lambda_{1}- \tilde x\vee\lambda_{2}\right)}\times \mathbf{1} \left(\lambda_{1}>  \tilde x\vee\lambda_{2}\right),
\end{eqnarray}
where  {$a \vee b = \max (a,b)$ and} $r$ is a rate function such that 
\begin{equation}\label{gamma}
r=\frac{1}{2}-\beta \gamma \int \frac{1}{\beta x-y}\, d\sigma_{\beta}(y)
-\frac{1-\gamma}{2x}
\end{equation}
with $\gamma=p/n$ and $\sigma_{\beta}$  denotes the probability distribution function of the  Marchenko--Pastur law such that 
\begin{equation}\label{sigmabeta}
\sigma_{\beta}(ds) = ({\beta \times 2\pi \gamma  s})^{-1} {\sqrt{(s-s_*)(s^*-s)}} \, \mathbf{1}(s\in [s_*, s^*])
ds
\end{equation}
with $s^*=\beta  (\sqrt{\gamma}+1  )^2 $ and  $s_*=\beta (\sqrt{\gamma}-1  )^2 $, and $\tilde x$ is a constant depending on $n$, $p$, $\beta$ and $x$ such that
 \begin{eqnarray*} \label{def:tildex}
 	\tilde x= { x\, \mathrm{tr} \,(\LL_{n-1,p-1,\beta}/n)}/{(p-x)}.
 \end{eqnarray*} 
 \item[Step\,3]  Based on the collected values $\lambda_1 \geq \cdots \geq \lambda_p$, a corresponding importance sampling estimate  can be computed as in \eqref{Lp} below and the value of the estimate is saved. 
\end{itemize}
 The three steps above are repeated  {at} every iteration. After the last iteration, the saved sampling estimates from all iterations are averaged to give an unbiased estimate of $\alpha_{n,p}(x)$. \end{algorithm}

 Now we  {detail} how  the importance sampling estimate \eqref{Lp}  {is computed at} every iteration of the algorithm.  Let $ Q$ be the measure induced by combining the above two-step sampling procedure. 
From \cite{dumitriu2002matrix}, under the change of measure $Q$, the density of $(\lambda^*_2,\ldots, \lambda^*_p) =n (\lambda_2,\ldots,\lambda_p)/(n-1)$ is
\begin{eqnarray*}\label{density1}
f^Q_{n,p}(\lambda_{2}^*,\ldots,\lambda_{p}^*)=
C_{n-1,p-1,\beta}\prod_{2\leq i<j\leq p}|\lambda^*_i-\lambda^*_j|^{\beta}
\times \prod_{i=2}^p (\lambda^*_i)^{ {\beta(n-p+1)}/{2}-1}\times 
e^{-(n-1) \sum_{i=2}^p \lambda^*_i/2}.
\end{eqnarray*}
This implies that the density function of $(\lambda_2,\ldots,\lambda_p)$ under $Q$ is
\begin{eqnarray}\label{density_l}
f^Q_{n,p}(\lambda_{2},\ldots,\lambda_{p})=
\left(\frac{n}{n-1}\right)^{ {\beta(n-1)(p-1)}/{2}}C_{n-1,p-1,\beta}\prod_{2\leq i<j\leq p}|\lambda_i-\lambda_j|^{\beta}
\times \prod_{i=2}^p \lambda_i^{ {\beta(n-p+1)}/{2}-1}\times
e^{-n\sum_{i=2}^p \lambda_i/2}.
\end{eqnarray}
Therefore $dQ/dP$ takes the form 
\begin{eqnarray*}
\frac{f^Q_{n,p}(\lambda_{2},\ldots,\lambda_{p})\times nr e^{-nr (\lambda_{1}-  \tilde x\vee\lambda_{2})}\times \mathbf{1}{(\lambda_{1} >  \tilde  x\vee \lambda_{2})}}{f_{n,p}(\lambda_1,\ldots,\lambda_{p})}=\frac{\left(\frac{n}{n-1}\right)^{{\beta(n-1)(p-1)}/{2}}C_{n-1,p-1,\beta}\,n r e^{-nr (\lambda_{1}-  \tilde x\vee\lambda_{2})}\times \mathbf{1}{(\lambda_{1} >  \tilde  x\vee \lambda_{2})}}{C_{n,p,\beta}  \prod_{i=2}^p(\lambda_{1}-\lambda_{i})
\times \lambda_{1}^{{\beta(n-p+1)}/{2}-1}\times
e^{-n \lambda_{1}/2}}.
\end{eqnarray*}
The corresponding importance sampling estimate  is given by
\begin{equation}\label{Lp}
 L_{n,p}(x)=
\mathbf{1}{(U_{n,p}>x)} \,  {dP}/{dQ} ,
\end{equation}
where $U_{n,p}$ is calculated with the sampled $\lambda_1,\ldots,\lambda_p$ based on Eq.~\eqref{def:u}. 

We claim that for the proposed Algorithm~\ref{algo:1}, with the choice of $r$  {specified} in \eqref{gamma}, 
 the importance sampling estimator $ L_{n,p}(x)$ is asymptotically efficient in estimating the target tail probability. This result is  {formally stated}  below  {and proved} in Section \ref{sec:proof}.  

\begin{theorem} \label{thm:mainthm}
When $p/n\to \gamma\in \mathbb{R}$, the estimator $L_{n,p}(x)$ in \eqref{Lp} is an asymptotically efficient estimator of $\alpha_{n,p}(x)$ for $x>(\sqrt{\gamma}+1 )^2$. 
\end{theorem}

\begin{remark}
\em
	Our discussion  {regarding} asymptotic efficiency focuses on the case of estimating rare-event tail probability  $\alpha_{n,p}(x)$, i.e., when $\{U_{n,p}>x\}$ corresponds to a rare event.  
 	When $x \leq (\sqrt{\gamma}+1 )^2$, $\{U_{n,p}>x\}$ is not  rare, and we can still apply the importance sampling algorithm with a reasonable positive $r$ value as the exponential distribution's rate. However, the theoretical properties of the importance sampling estimator  {must then be} studied under a different framework;  {this issue is} not pursued here. 
	\end{remark}

\begin{remark}\label{remark:constantdiff}
\em
 We explain the Marchenko--Pastur form of (\ref{sigmabeta}). When the entries of $\mathbf{X}$  have mean 0 and variance 1 ($\beta=1$ and $2$), the  Marchenko--Pastur law for the eigenvalues of $\mathbf{X}^{H}\mathbf{X} /n$ takes the standard form 
 \begin{eqnarray}\label{MPlaw2}
		f(d\bar{s})= ({2\pi \gamma \bar{s}})^{-1} {\sqrt{( \bar{s}_{+}-\bar{s})(\bar{s}-\bar{s}_{-})}} \, \mathbf{1}( {\bar s}\in [\bar{s}_{-},\bar{s}_{+} ])d\bar{s}
	\end{eqnarray} 
	with  $\bar{s}_{-}=(1-\sqrt{\gamma})^2$ and $\bar{s}_{+}=(1+\sqrt{\gamma})^2$; see, e.g., Theorem~3.2 in~\citep{paul2014random}. For the setting considered of this paper, the real case ($\beta=1$) has $\sigma^2= 1$, so \eqref{sigmabeta} and \eqref{MPlaw2} are consistent. In contrast,  the complex case ($\beta=2$) has $\sigma^2= 2$ and therefore   \eqref{sigmabeta} and \eqref{MPlaw2} are different up to a factor of $\beta=2$.
Specifically, let  $(\bar{\lambda}_1, \ldots ,\bar{\lambda}_p)$ and $(\lambda_1,\ldots, \lambda_p)$  be eigenvalues of $ \mathbf{X}^{H}\mathbf{X}/n$ when $\mathbf{X}$ has iid  entries of $\mathcal{CN}(0,1)$ and $\mathcal{CN}(0,2)$, respectively. Then we  know that $(\lambda_{1},\ldots, \lambda_{p})\sim 2(\bar{\lambda}_{1},\ldots, \bar{\lambda}_{p})$ and \eqref{MPlaw2} implies the empirical distribution in (\ref{sigmabeta}). 
\end{remark}

\begin{remark} \label{rm:compare}
\em
	 We discuss the differences between the proposed method and the method in \cite{xu2016rare}  on the largest eigenvalue, which also employs an importance sampling technique.
First, the two methods have different targets, i.e., $\Pr(\lambda_1> x)$ in \cite{xu2016rare} and $\Pr( U_{n,p}>x)$ here, and therefore use different  {changes} of measure to construct efficient importance sampling estimators. As discussed in Section~\ref{description}, in order to achieve  {asymptotic} efficiency,    the change of measures should approximate the target conditional distribution measures, i.e., $\Pr(~\cdot \mid  \lambda_1> x)$ in \cite{xu2016rare} and $\Pr(~\cdot \mid U_{n,p}>x)$ in this paper. Due to the difference between the two conditional distributions, different  {changes} of measure are constructed in the two methods. Specifically,  {Jiang et al.} \cite{xu2016rare} sample the largest eigenvalue $\lambda_1$ from a truncated  exponential distribution depending on the second largest eigenvalue $\lambda_2$, while the present work samples $\lambda_1$ from an exponential distribution depending on eigenvalues $\lambda_2,\ldots,\lambda_p$. 
Second, the proof techniques of the main asymptotic results in the two papers are also different. In particular, to show the asymptotic efficiency of the importance sampling estimators as defined in \eqref{logeff}, we need to derive asymptotic approximations for both the rare-event probability $\alpha_{n,p}(x)$ and the second moments of the importance sampling estimator ${\rm E}_Q\{L_{n,p}^2(x)\}$. Even though the largest eigenvalue $\lambda_1$ and the ratio statistic $U_{n,p}$ have similar large deviation approximation results for their tail probabilities, the asymptotic approximations for the second moments of the importance sampling estimators are different   due to the differences between the considered  {changes} of measure as well as the effect of the trace term in $U_{n,p}$. Please refer to the proof for more details.
\end{remark}

\begin{remark} \label{rm:pnswitch}
\em
The method and the theoretical results can be easily extended  from the case $p \leq n$ to the case $p\geq n$ by switching the labels of $n$ and $p$ and changing $\gamma$  to $1/\gamma $ correspondingly. Note that when $p\geq n$, the eigenvalues of $\mathbf{X}^H \mathbf{X}/n$ and $ \mathbf{X}\mathbf{X}^H/p$ give the same test statistic $U_{n,p}$ as defined in \eqref{def:u}, which is because  $\mathbf{X}^H \mathbf{X}$ and $\mathbf{X}\mathbf{X}^H$ have the same set of  nonzero eigenvalues and $U_{n,p}$ is scale invariant.
By symmetry, when $p\geq n$, the joint density function of the eigenvalues of $\XX \XX^H/p$  have the same form  as  \eqref{originaldensity},  except that the labels of $n$ and $p$ are switched. Therefore, the cases when $p\leq n$ and $p\geq n$ are equivalent up to the label switching.  Note that after  $p/n$ is changed to $n/p$,  $\gamma$ becomes $1/\gamma$ correspondingly.
\end{remark}

\section{Numerical study}\label{sec:num}

We conducted numerical studies to evaluate the performance of our algorithm. We first took combinations $(n,p) \in \{ (100,10) , (100,20)$, $(500,20)$, $(1000,50)\}$, and   $\beta=1$ and $2$, respectively. 
Then we compared our algorithm with other methods and present the results in  Table \ref{testre1} and \ref{testre2}.

For the proposed importance sampling estimator, we repeated $N_{IS}=10^4$ times and show the estimated probabilities (``$EST_{IS}$" column) along with the estimated standard deviations  of $L_{n,p}$, i.e., $\sqrt{Var^Q(L_{n,p})}$  (``$SD_{IS}$" column).  
The ratios between estimated standard deviations and estimates (``$SD_{IS}/EST_{IS}$" column)  {reflect} the efficiency of the algorithms. 
Note that with $N_{IS}=10^4$ replications, the standard error of the estimate  is $SD_{IS}/\sqrt{N_{IS}}=SD_{IS}/100$. 

In addition, three alternative methods were considered,  {namely} the direct Monte Carlo, the  {Tracy--Widom} distribution approximation, and the corrected  {Tracy--Widom} approximation \citep{nadler2011distribution}. We computed direct Monte Carlo estimates (``$EST_{DMC}$" column) with $N_{DMC}=10^6$ independent replications. We present the standard deviation of direct Monte Carlo estimates (``$SD_{DMC}$" column) and the ratios between estimated standard deviations and estimates (``$SD_{DMC}/EST_{DMC}$"). 
In addition, we used the approximation of  {Tracy--Widom} distribution (``$TW$" column) specified in Eq.~\eqref{tw}.  The $TW(x)$ is computed from  {the} \texttt{RMTstat} package in \textsf{R}.
Furthermore, following \cite{nadler2011distribution}, we computed the  {Tracy--Widom} approximation with correction term (``$c.TW$" column), viz.
\begin{eqnarray} \label{corTW}
 \Pr\left ( \frac{U- \mu_{n,p}}{\sigma_{n,p}}  > x \right) \approx  1- \mathcal{TW}_{\beta}(x) + \frac{1}{2} \left (  \frac{2}{np} \right) \left ( \frac{\mu_{n,p}}{\sigma_{n,p}}   \right)^2 \mathcal{TW}_{\beta}^{''} (x),
\end{eqnarray}
where   $\mathcal{TW}^{''} (x)$ is computed numerically via a standard central differencing scheme with $\Delta x = 10^{-3}$. When $\beta=1$, $\mu$ and $\sigma$ is chosen according to Eq.~(\ref{parameter}). When $\beta=2$, $\mu$ and $\sigma$ is chosen according to \cite{karouicomplexrate}.
 
We can see from Tables~\ref{testre1} and~\ref{testre2} that 
the  {Tracy--Widom} distribution (``$TW$" column)  significantly overestimates the tail probabilities for all considered settings and the finding is consistent with that in  \cite{nadler2011distribution}.   {Furthermore,} the corrected  {Tracy--Widom} approximation (``$c.TW$" column) underestimates the tail probability $\alpha_{n,p}(x)$ and goes to a negative number as  $\alpha_{n,p}(x)$  {becomes} small. 

Since the proposed importance sampling and the direct Monte Carlo method are both unbiased estimators, next we compare their computational efficiency. 
As discussed in Section~\ref{description}, for the average estimator $\bar{L}_{n,p}(x)= \{L_{n,p}^{(1)}(x) + \cdots + L_{n,p}^{(N)}(x)\}/N$, 
``$SD_{IS}/EST_{IS}$" and ``$SD_{DMC}/EST_{DMC}$" can be used as a  measure of the computational efficiency in terms of iteration numbers.
 From the results in Tables \ref{testre1} and \ref{testre2},  as $\alpha_{n,p}(x)$ decreases, ``$SD_{DMC}/EST_{DMC}$"  {grows} quickly and even becomes not available. In contrast, ``$SD_{IS}/EST_{IS}$" increases slowly and is generally smaller than ``$SD_{DMC}/EST_{DMC}$",  showing that the proposed importance sampling is  more efficient than the direct Monte Carlo method.  

\begin{table}[!htbp]
\centering
\caption{Estimation Results for $\beta=1$.} \label{testre1}
{\small
\subcaption{$n=100 $, $p=10$}
\label{my-label}
\begin{tabular}{ccccccccc}
$x$   & $\mathrm{EST_{IS}}$       & $\mathrm{SD_{IS}}$       & $\mathrm{SD_{IS}/EST_{IS}}$ & $\mathrm{EST_{DMC}}$ & $\mathrm{SD_{DMC}}$ & $\mathrm{SD_{DMC}/EST_{DMC}}$ & $\mathrm{c.TW}$    &  $\mathrm{TW}$    \\ \hline 
1.80 & 2.44e-2    & 1.25e-1     & 5.14    &  2.46e-2     & 1.55e-1      & 6.30       & 2.58e-2   & 5.07e-2   \\  
1.95 &1.02e-3 & 5.00e-3 & 4.89 & 1.08e-3 &  3.28e-2 &30.46 & 3.90e-4 & 4.37e-3 \\  
1.98 & 5.32e-4 & 3.55e-3 &  6.66 & 5.57e-4 & 2.36e-2 &  42.36 & 4.96e-6  & 2.48e-3 \\  
2.10 & 2.43e-5 & 2.48e-4 & 10.22   & 2.20e-5  & 4.69e-3  & 213.20   & --7.46e-5& 2.07e-4 \\ 
2.30 & 5.25e-8 & 7.72e-7 & 14.71   & 0           & 0           & NaN           & 0   &  0      \\ \hline
\end{tabular}

\vspace{1em}

\subcaption{$n=100 $, $p=20$}
\begin{tabular}{ccccccccc}
$x$   & $\mathrm{EST_{IS}}$       & $\mathrm{SD_{IS}}$       & $\mathrm{SD_{IS}/EST_{IS}}$ & $\mathrm{EST_{DMC}}$ & $\mathrm{SD_{DMC}}$ & $\mathrm{SD_{DMC}/EST_{DMC}}$ & $\mathrm{c.TW}$    &  $\mathrm{TW}$    \\ \hline 
2.10 & 9.14e-2 & 3.73e-1 & 4.09 & 8.99e-2 & 2.86e-1 &3.18 & 9.29e-2 & 1.21e-1 \\  
2.30 & 2.86e-3 & 2.04e-2 & 7.13 & 2.71e-3  & 5.20e-2 & 19.19 & 2.31e-3 &   6.09e-3 \\  
2.40 & 3.44e-4 & 2.60e-3 & 7.54 & 3.11e-4 & 1.76e-2&  56.70 & 1.54e-4 & 9.07e-4  \\ 
2.50 & 2.89e-5& 2.01e-4 & 6.95 & 2.60e-5 & 5.10e-3 & 196.11 &  --6.13e-6 & 1.05e-4 \\  
2.70 & 1.50e-7 &  1.78e-6 & 11.85 & 0&0&NaN& 0& 0\\ \hline
\end{tabular}
\vspace{1em}
\subcaption{$n=500 $, $p=20$}

\begin{tabular}{ccccccccc}
$x$   & $\mathrm{EST_{IS}}$       & $\mathrm{SD_{IS}}$       & $\mathrm{SD_{IS}/EST_{IS}}$ & $\mathrm{EST_{DMC}}$ & $\mathrm{SD_{DMC}}$ & $\mathrm{SD_{DMC}/EST_{DMC}}$ & $\mathrm{c.TW}$    &  $\mathrm{TW}$    \\ \hline 
1.46 & 4.64e-2 & 2.21e-1 & 4.76 & 4.68e-2  &  2.11e-1 & 4.51 &4.87e-2 & 6.51e-2 \\  
1.51 & 3.98e-3& 2.16e-2&5.43&3.70e-3& 6.07e-2 &16.40& 3.70e-3 & 7.03e-3  \\  
1.56 & 1.57e-4 & 7.13e-4 & 4.54 &1.55e-4 & 1.24e-2 &80.32& 1.28e-4& 4.40e-4  \\ 
1.62& 2.14e-6 &1.49e-5 & 6.97 & 3.00e-6 &1.73e-3&577.35 &--1.87e-6 &6.71e-6  \\  
1.70&2.43e-9& 2.72e-8 &11.20 & 0&0&NaN& 0& 0 \\ \hline
\end{tabular}

\vspace{1em}
\subcaption{$n=1000 $, $p=50$}
\begin{tabular}{ccccccccc}
$x$   & $\mathrm{EST_{IS}}$       & $\mathrm{SD_{IS}}$       & $\mathrm{SD_{IS}/EST_{IS}}$ & $\mathrm{EST_{DMC}}$ & $\mathrm{SD_{DMC}}$ & $\mathrm{SD_{DMC}/EST_{DMC}}$ & $\mathrm{c.TW}$    &  $\mathrm{TW}$    \\ \hline 
1.52 &  2.75e-2 & 1.29e-1&4.70  & 2.90e-2 &  1.68e-1 &  5.78 &  2.96e-2 & 3.59e-2 \\  
1.55 & 2.51e-3 & 1.16e-2& 4.63 &2.57e-3& 5.06e-2& 19.71 &  2.53e-3  & 7.98e-4 \\ 
1.60 &1.41e-5&  5.25e-5 & 3.72&  2.20e-5  &4.69e-3&  213.20 & 1.15e-5& 3.25e-5 \\  
1.62 & 1.40e-6 & 8.70e-6 & 6.21 & 2.00e-6 &1.41e-3&707.11& --7.93e-7& 6.71e-6 \\  
1.66 & 7.49e-9 & 3.69e-8 & 4.93 & 0&0&NaN&0& 0 \\ \hline
\end{tabular}
}
\end{table}

\begin{table}[!htbp]
\centering
\caption{Estimation Results for for $\beta=2$.} \label{testre2}
\subcaption{$n=100 $, $p=10$}

\label{my-label}
{\small
\begin{tabular}{ccccccccc}
$x$   & $\mathrm{EST_{IS}}$       & $\mathrm{SD_{IS}}$       & $\mathrm{SD_{IS}/EST_{IS}}$ & $\mathrm{EST_{DMC}}$ & $\mathrm{SD_{DMC}}$ & $\mathrm{SD_{DMC}/EST_{DMC}}$ & $\mathrm{c.TW}$    &  $\mathrm{TW}$    \\ \hline  
1.77  & 3.72e-3& 3.34e-2  & 8.98 & 3.79e-3 & 6.15e-2 &16.21  & 2.20e-3 & 1.26e-2  \\  
1.81  & 9.21e-4 & 1.32e-2 & 14.34 & 8.97e-4 & 2.99e-2 & 33.37  &-1.36e-4    & 4.42e-3  \\  
1.91 & 1.89e-5  & 3.28e-4 & 17.37 &1.70e-5 & 4.12e-3 & 242.53 & -1.22e-4 & 2.11e-4  \\  
1.93 & 6.68e-6& 8.44e-5 & 12.64 &4.00e-6  & 2.00e-3  & 500   & -7.44e-5   & 1.07e-4  \\  
1.99 & 2.98e-7& 4.25e-6 & 14.27& 0  & 0 &NaN  & -1.29e-5   &1.24e-5   \\ \hline
\end{tabular} 

\vspace{1em}

\subcaption{$n=100 $, $p=20$}
\begin{tabular}{ccccccccc}
$x$   & $\mathrm{EST_{IS}}$       & $\mathrm{SD_{IS}}$       & $\mathrm{SD_{IS}/EST_{IS}}$ & $\mathrm{EST_{DMC}}$ & $\mathrm{SD_{DMC}}$ & $\mathrm{SD_{DMC}/EST_{DMC}}$ & $\mathrm{c.TW}$    &  $\mathrm{TW}$    \\ \hline  
2.10 &1.20e-2 & 7.99e-2 &6.68&1.45e-2  &1.20e-1   & 8.23& 1.41e-2    &2.70e-2   \\ 
2.18 &1.04e-3  & 7.59e-3 & 7.28 & 1.34e-3 & 3.66e-2  & 27.29 & 8.64e-4   & 3.65e-3  \\  
2.30  & 2.18e-5 & 3.47e-4 & 15.94 & 2.30e-5 & 4.80e-3 & 208.51 &  -2.06e-5  & 8.86e-5  \\ 
2.38 & 6.73e-7 & 1.94e-5 & 28.86  &  1.00e-6 & 1.00e-3 & 1000  &  -2.70e-6  &4.83e-6  \\  
2.46 & 1.63e-8 & 2.83e-7  &  17.36 & 0 & 0 & NaN &  -1.73e-7  & 1.93e-7  \\ \hline  
\end{tabular}
\vspace{1em}
\subcaption{$n=500 $, $p=20$}
\begin{tabular}{ccccccccc}
$x$   & $\mathrm{EST_{IS}}$       & $\mathrm{SD_{IS}}$       & $\mathrm{SD_{IS}/EST_{IS}}$ & $\mathrm{EST_{DMC}}$ & $\mathrm{SD_{DMC}}$ & $\mathrm{SD_{DMC}/EST_{DMC}}$ & $\mathrm{c.TW}$    &  $\mathrm{TW}$    \\ \hline 
1.45 & 8.04e-3 &5.49e-4&6.84&8.98e-3 &9.43e-2 &10.51& 8.95e-3 &1.58e-2   \\  
1.48& 6.56e-4 &8.02e-3&12.22& 6.49e-4 & 2.55e-2 &39.24  &    5.07e-4&1.59e-3  \\ 
1.50 & 8.77e-5 & 1.16e-3 & 13.18& 8.60e-5  &9.27e-3 & 107.83 & 3.88e-5& 2.70e-4  \\  
1.525 & 5.05e-6 &5.37e-5& 10.63 &  8.00e-6&2.83e-3 & 353.55  &-1.87e-6 & 2.28e-5  \\  
1.55 & 1.85e-7 & 1.71e-6 & 9.28& 0 &0  & NaN& -4.66e-7 & 1.49e-6   \\ \hline
\end{tabular}

\vspace{1em}
\subcaption{$n=1000 $, $p=50$}
\begin{tabular}{ccccccccc}
$x$   & $\mathrm{EST_{IS}}$       & $\mathrm{SD_{IS}}$       & $\mathrm{SD_{IS}/EST_{IS}}$ & $\mathrm{EST_{DMC}}$ & $\mathrm{SD_{DMC}}$ & $\mathrm{SD_{DMC}/EST_{DMC}}$ & $\mathrm{c.TW}$    &  $\mathrm{TW}$    \\ \hline  
1.51 & 5.85e-3& 6.67e-2&  11.39 &5.20e-3&7.19e-2&13.83&5.31e-3&7.46e-3 \\  
1.53 &2.65e-4&1.96e-3&  7.39 &3.04e-4& 1.74e-2 & 57.35 &2.98e-4& 5.32e-4 \\
1.56 & 1.72e-6& 1.84e-5& 10.72&0  & 0 &NaN & 1.33e-6  & 4.20e-6  \\  
1.58&3.15e-8&2.86e-7& 9.10&0&0&NaN&1.46e-8& 9.85e-8 \\  
1.60 &4.24e-10& 3.80e-9&8.97&0&0&NaN&-6.21e-11& 1.56e-9 \\ \hline
\end{tabular}}

\end{table}

 {As a further illustration}, we  compared the iteration numbers $N_{IS}$ and $N_{DMC}$ that would be needed to achieve the same level of relative standard errors of the estimators. Specifically, in order to have the same  ratios of  the  standard errors  to the estimates, i.e., $SE_{IS}/EST_{IS}= (SD_{IS}/\sqrt{N_{IS}})/EST_{IS}$ and $SE_{DMC}/EST_{DMC}=(SD_{DMC}/\sqrt{N_{DMC}})/EST_{DMC}$,  obtained under the importance sampling and direct direct Monte Carlo, respectively, we need 
\begin{equation}\label{number}
\frac{N_{DMC}}{N_{IS}} = \frac{ (SD_{DMC}/EST_{DMC})^2}{(SD_{IS}/EST_{IS})^2}.	
\end{equation}
Based on the above equation,  the simulation results show that to have a similar standard error obtained under the importance sampling,  the direct Monte Carlo method needs more iterations  as $\alpha_{n,p}(x)$ goes small. 
For example, from Table~1,  when $n=100$, $p=10$ and $x=2.1$,    
we need  $N_{DMC}$ to be approximately $4.3 \times 10^{2}$ times larger than $N_{IS}$;
when $n=1000$, $p=50$ and $x=1.62$, we need $N_{DMC}$  {to be} about $1.3\times 10^4$ 
times larger.

 Besides the iteration numbers, we compared the average time cost of each iteration  under the importance sampling and the direct Monte Carlo method,  respectively. For the direct Monte Carlo, two methods were considered in computing the eigenvalues. 
The first method directly  computes the test statistic  $U_{n,p}$  using the eigen-decomposition of a randomly sampled Wishart matrix. 
The second method  computes the eigenvalues from the tridiagonal representation form as in Step~1 of Algorithm~1. 
We ran $10^4$ iterations for all the methods and report the average  time of one iteration in Table \ref{timeres}, where  the first  method of the direct Monte Carlo is denoted as $T_{DMC\_1}$, the second method is denoted as   $T_{DMC\_2}$, and the importance sampling method is denoted as $T_{IS}$. 
The simulation results show  that  $T_{DMC\_1}$ has the highest time cost per iteration, while  $T_{DMC\_2}$ and $T_{IS}$ are similar. 

We further explain the simulation results from the perspective of algorithm complexity. 
For each iteration, the first direct Monte Carlo method samples a $p\times p$ Wishart matrix and performs its eigen-decomposition,  {whose cost is typically of the order of} $O(p^3)$.
The second direct Monte Carlo method and the importance sampling only need to sample $O(p)$ number of chi-square random variables and then decompose a symmetric tridiagonal matrix,  {at a cost of $O(p^2)$} per iteration \cite{Demmel:1997:ANL:264989}. Although the importance sampling also samples from an exponential distribution in Step~2, the distribution parameters can be calculated in advance and it does not affect the overall complexity much. 
Therefore,  the time complexity of the algorithm $T_{DMC\_1}$ is higher while  $T_{DMC\_2}$ and $T_{IS}$ are similar per iteration.
Together with the result in \eqref{number},  we can see that the importance sampling is more efficient than the direct Monte Carlo method in terms of both the iteration number and the overall time cost. 

To further check the influence of replication number $N_{IS}$ of the importance sampling algorithm, we focus on the case $(n, p) = (100, 10)$ and compare the performance of different $N_{IS}$s.
In order to obtain accurate reference values of the tail probabilities, we used direct Monte Carlo with repeating time $N_{DMC}=10^8$ to estimate multiple tail probabilities $\alpha_{n,p}(x)$s ranging from $10^{-2}$ to $10^{-6}$ under $\beta = 1$ and $\beta = 2 $, respectively. 
Then we estimated the corresponding $\alpha_{n,p}(x)$s using our algorithm with $N_{IS} = 10^4 , 10^5 , 10^6$, respectively. 

\begin{figure}[htbp]
\begin{subfigure}{.5\textwidth}
  \centering
  \includegraphics[width=.9\linewidth]{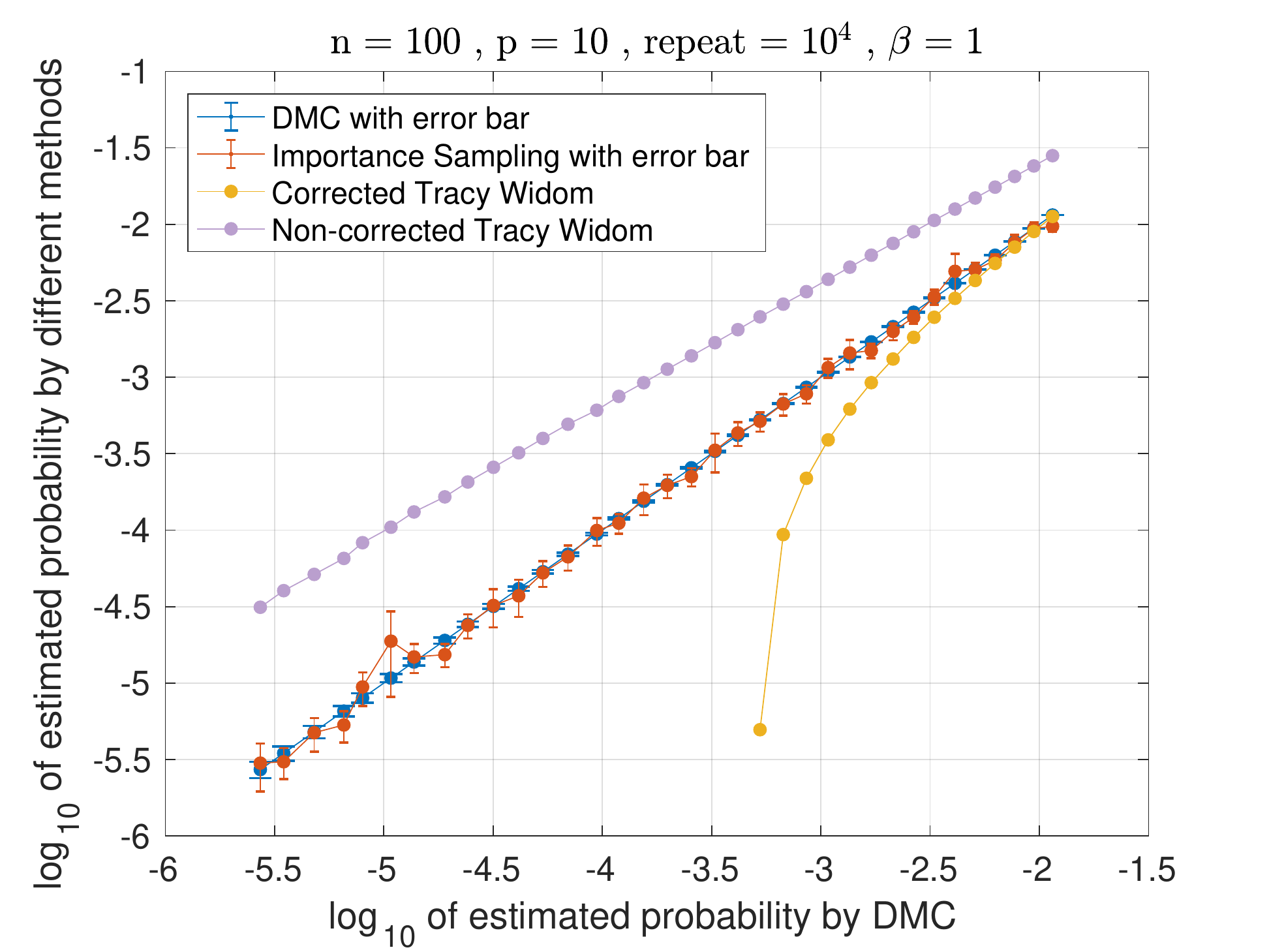}
  \caption{}
  \label{fig:ims4}
\end{subfigure}%
\begin{subfigure}{.5\textwidth}
  \centering
 \includegraphics[width=.9\linewidth]{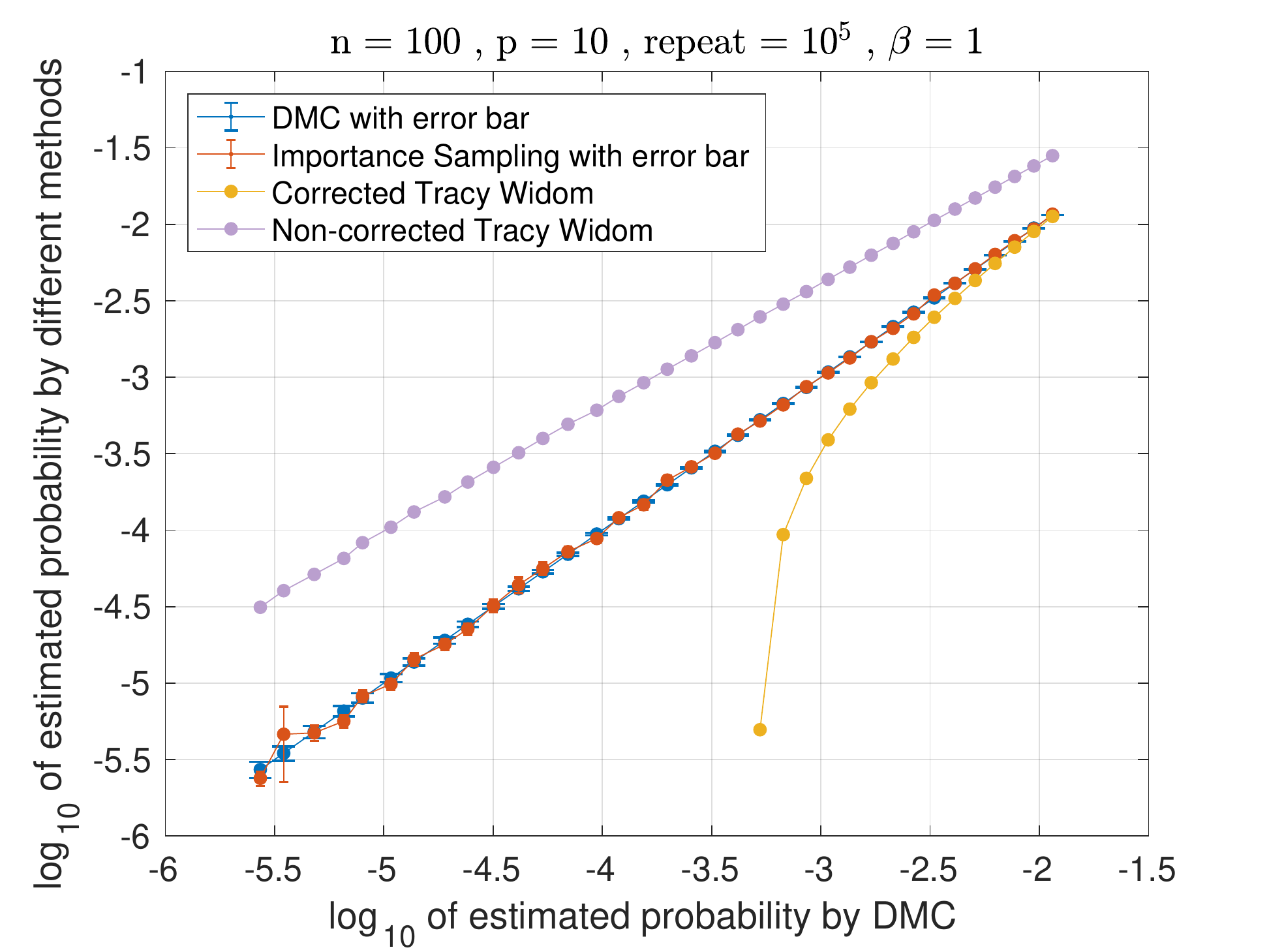}
 \caption{}
  \label{fig:ims5}
\end{subfigure}

\begin{subfigure}{.5\textwidth}
  \centering
  \includegraphics[width=.9 \linewidth]{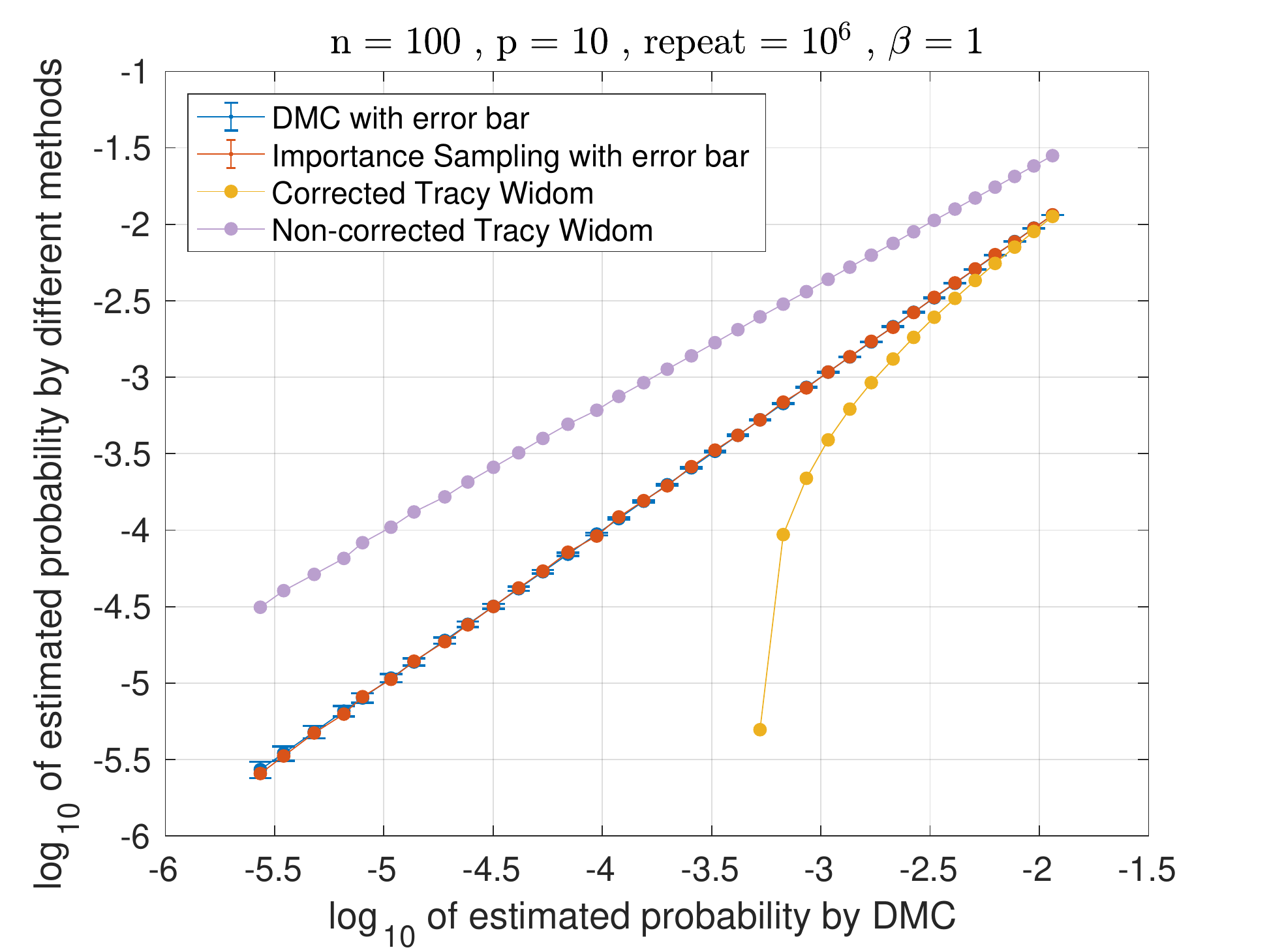}
\caption{}
  \label{fig:ims6}
\end{subfigure}%
\begin{subfigure}{.5\textwidth}
  \centering
  \includegraphics[width=.9\linewidth]{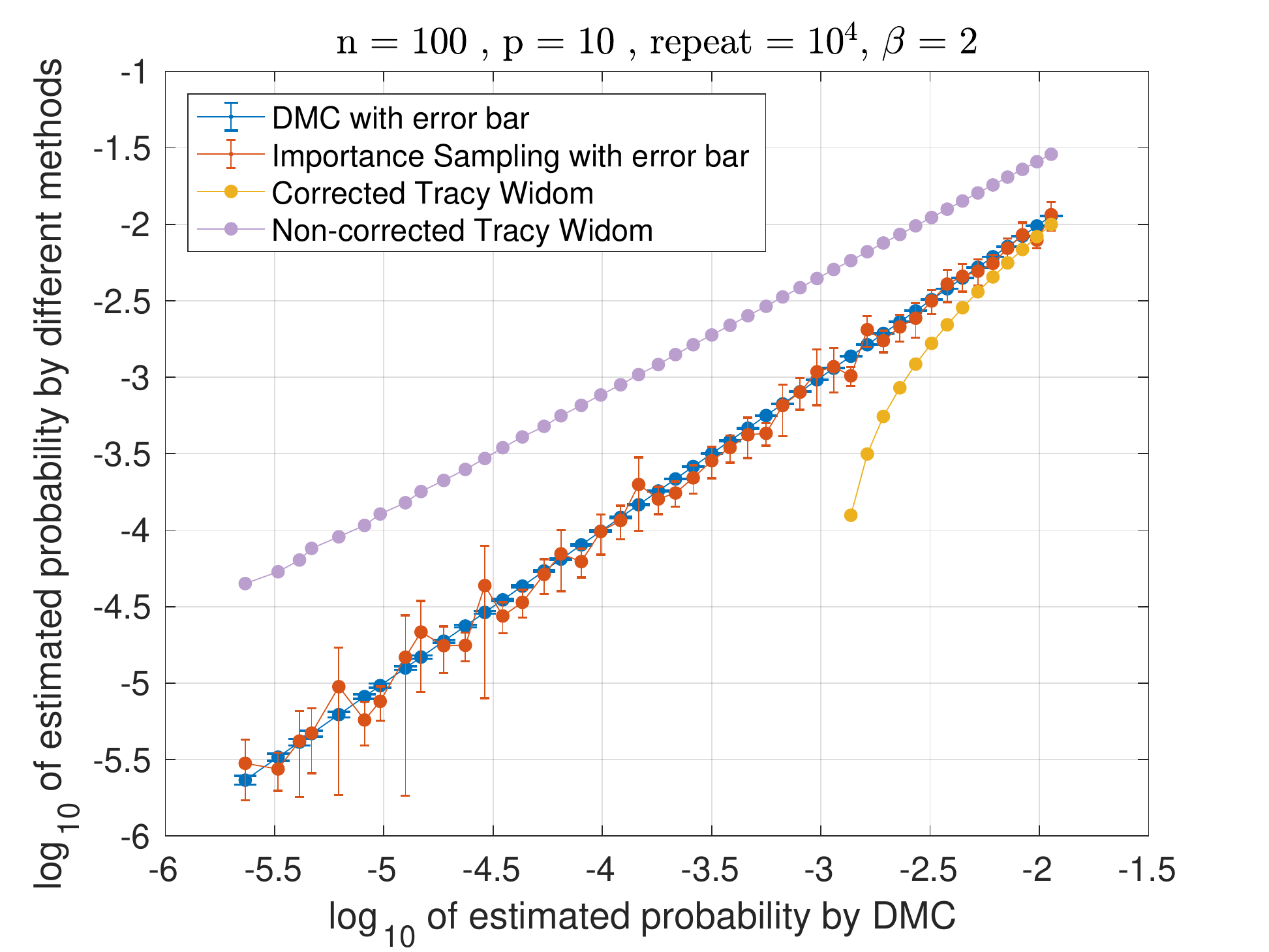}
  \caption{}
  \label{fig:com4}
\end{subfigure}
\begin{subfigure}{.5\textwidth}
  \centering
  \includegraphics[width=.9\linewidth]{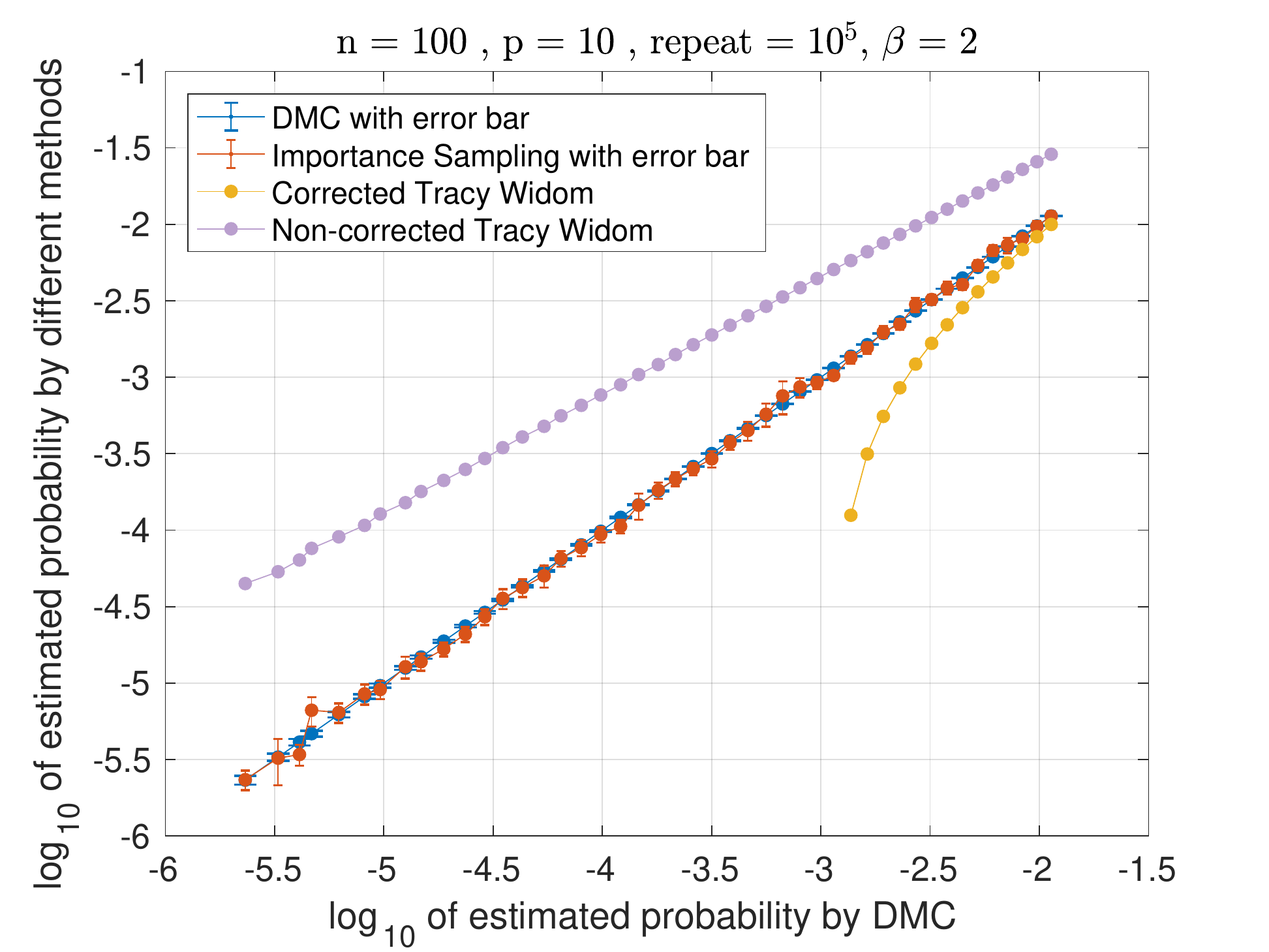}
  \caption{}
  \label{fig:com5}
\end{subfigure}%
\begin{subfigure}{.5\textwidth}
  \centering
 \includegraphics[width=.9\linewidth]{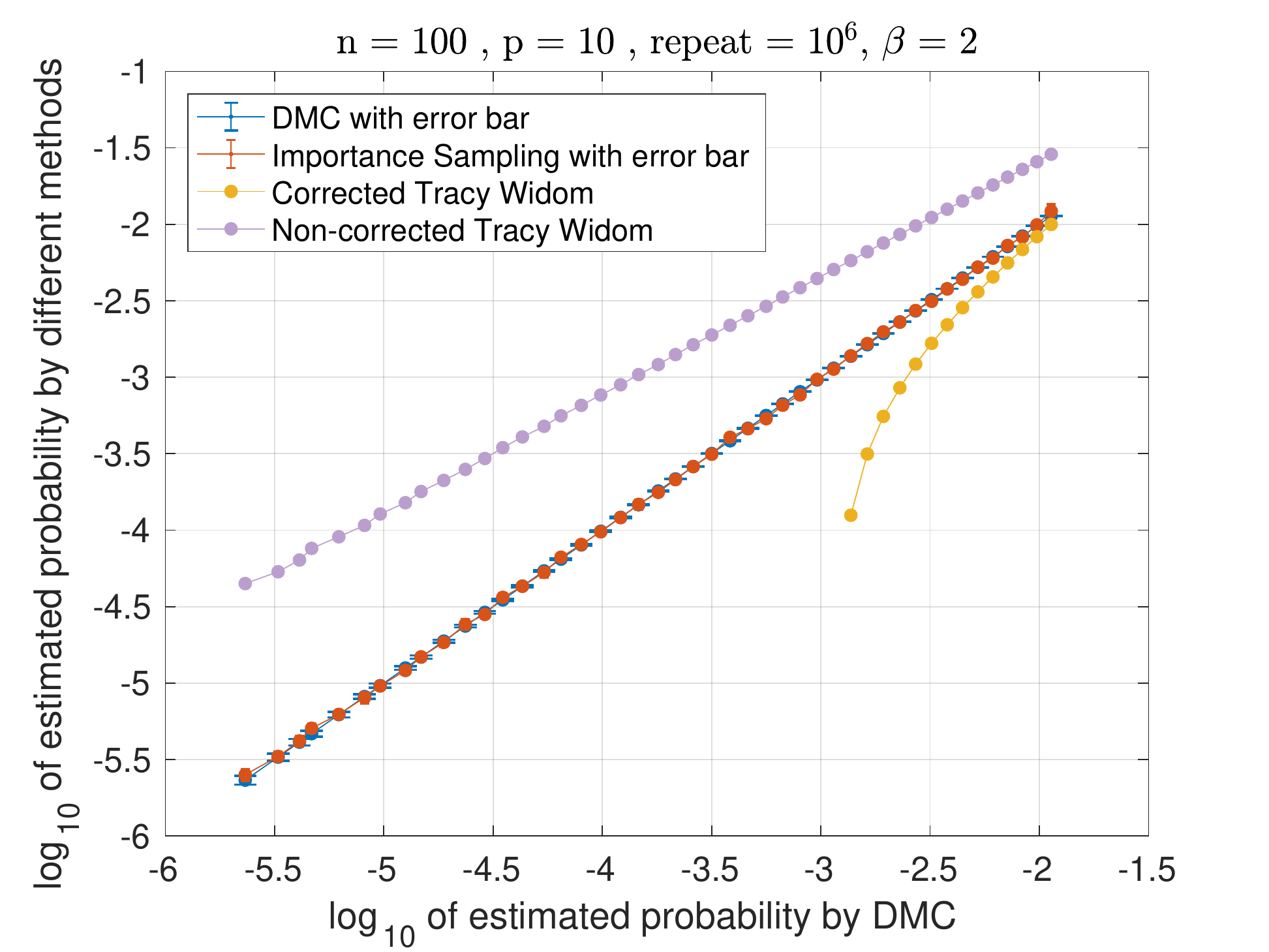}
 \caption{}
  \label{fig:com6}
\end{subfigure}
\caption{Estimation results for $n=100$ and $p=10$.}
\label{fig:plotres}
\end{figure}

The results are presented in Figure~\ref{fig:plotres}, where the $x$-axis represents the reference values $\log_{10} (EST_{DMC} )$. The line ``\textit{DMC with error bar}" represents the (approximated) pointwise $95\%$ confidence intervals, viz.
$$
[ \log_{10}  (EST_{DMC}-2\times SD_{DMC}/\sqrt{N_{DMC}}  ),  \log_{10}   (EST_{DMC}+2\times SD_{DMC}/ \sqrt{N_{DMC}}  )   ].
$$
Similarly, the line ``\textit{Importance Sampling with error bar}" represents the importance sampling estimates and pointwise $95\%$ confidence intervals, viz.
$$ 
[ \log_{10}  ( EST_{IS}-2\times SD_{IS}/\sqrt{N_{IS} }   ),\log_{10}  (EST_{IS} + 2\times SD_{IS}/\sqrt{N_{IS}} ) ].
$$ 

 {One can surmise from the figures that} the proposed algorithm  {gives reliable estimates of probabilities} as small as $10^{-6}$   with $N_{IS}=10^4$, which is   more efficient than directed Monte Carlo and   more accurate than  {Tracy--Widom} approximations. Furthermore,  Figure~\ref{fig:plotres} shows that the   algorithm improves as the number of iterations increases.  We also plot the  {Tracy--Widom} approximations in \eqref{tw} and \eqref{corTW} in Figure~\ref{fig:plotres} for comparison.

 Figure~\ref{fig:plotres} shows that without  correction, the  {Tracy--Widom} distribution in \eqref{tw}  is not accurate and overestimates the probabilies. The correction term in \eqref{corTW} improves the approximation when the probability is larger than the scale of about $10^{-2}$, which is consistent with the result in \cite{nadler2011distribution}. But when the   probability  {gets} smaller, the corrected approximation has larger deviation  from true values (on the $\log_{10}$ scale) and even becomes negative. 
Note that since we cannot plot the $\log_{10}$ of negative numbers in the figures,   the lines of the corrected  {Tracy--Widom} approximations appear to be shorter. These results validate the results in Table \ref{testre1} and \ref{testre2}.

\begin{table}[!htbp]
\centering
\caption{{Estimation of Time.}} \label{timeres}
\subcaption{\ $\beta=1$} \label{time-label1}
{\small
\begin{tabular}{cccccc}
$n$ & $p$ & $x$ & $T_{DMC\_1}$ & $T_{DMC\_2}$ & $T_{IS}$ \\ \hline  
100 & 10 & 1.95 & 1.28e-03 & 7.26e-04 & 8.89e-05  \\  
100 & 10 & 1.98 & 1.15e-03 & 9.23e-05 & 8.51e-05 \\  
100 & 20 & 2.3 & 1.58e-03 & 7.35e-05 & 6.84e-05 \\  
100 & 20 & 2.4 & 1.65e-03 & 1.79e-04 & 6.33e-05 \\  
500 & 20 & 1.51 & 1.27e-03 & 9.87e-05 & 9.32e-05 \\  
500 & 20 &  1.56 & 1.67e-03 & 7.39e-05 & 8.82e-05 \\  
1000 & 50 & 1.55 & 3.19e-03 & 1.05e-04 & 1.56e-04 \\  
1000 & 50 & 1.6 & 3.12e-03 & 9.76e-05 & 1.34e-04 \\ \hline
 \end{tabular} 
\vspace{1em}

\subcaption{\ $\beta=2$} \label{time-label2}
 
\begin{tabular}{cccccc}
$n$ & $p$ & $x$ & $T_{DMC\_1}$ & $T_{DMC\_2}$ & $T_{IS}$ \\ \hline  
100 & 10 & 1.77 & 1.87e-03 & 1.75e-04 & 6.08e-05 \\  
100 & 10 & 1.81 & 1.85e-03 & 5.47e-05 & 5.90e-05 \\  
100 & 20 & 2.18 & 2.86e-03 & 8.37e-05 & 1.20e-04 \\  
100 & 20 & 2.3 & 2.69e-03 & 1.11e-04 & 6.69e-05 \\  
500 & 20 & 1.45 & 2.79e-03 & 8.46e-05 & 7.01e-05 \\  
500 & 20 & 1.48 & 3.53e-03 & 7.24e-05 & 8.90e-05 \\  
1000 & 50 & 1.53 & 8.65e-03 & 9.03e-05 & 1.53e-04 \\  
1000 & 50 & 1.56 & 8.35e-03 & 9.61e-05 & 1.49e-04 \\ \hline
  \end{tabular}}
\end{table}

\begin{table}[!htbp]
\centering
\caption{$U_{n,p}^k$ Results for $\beta=1$.} \label{testrek1}

\subcaption{$n=100 $, $p=50$, $k=2$}
\label{my-label}
\begin{tabular}{ccccccccc}
$x$   & $\mathrm{EST_{IS}}$       & $\mathrm{SD_{IS}}$       & $\mathrm{SD_{IS}/EST_{IS}}$ & $\mathrm{EST_{DMC}}$ & $\mathrm{SD_{DMC}}$ & $\mathrm{SD_{DMC}/EST_{DMC}}$    \\ \hline  
5.9 & 1.14e-03 & 6.70e-03 & 5.89 & 1.55e-03 & 3.93e-02  & 25.41  \\  
6.0 & 3.22e-04 & 3.80e-03 &   11.78 & 2.98e-04 & 1.73e-02 &  57.92 \\  
6.1 & 5.68e-05 & 9.37e-04 & 16.49  &5.50e-05 &7.42e-03 & 134.84 \\  
6.4 & 1.09e-07 & 3.21e-06 & 29.50    & 0           & 0           & NaN     \\ \hline
\end{tabular}

\vspace{1em}

\subcaption{$n=100 $, $p=50$, $k=3$}
\label{my-label}
\begin{tabular}{ccccccccc}
$x$   & $\mathrm{EST_{IS}}$       & $\mathrm{SD_{IS}}$       & $\mathrm{SD_{IS}/EST_{IS}}$ & $\mathrm{EST_{DMC}}$ & $\mathrm{SD_{DMC}}$ & $\mathrm{SD_{DMC}/EST_{DMC}}$    \\ \hline  
8.4 &1.56e-03 & 1.77e-02 & 11.36 & 1.55e-03 & 3.93e-02  & 25.41  \\  
8.5 &  4.22e-04 & 5.48e-03 &  12.98 & 4.04e-04 & 2.01e-02 & 49.74  \\  
8.7 & 1.46e-05 & 2.99e-04 & 20.44   & 1.80e-05  & 4.24e-03 &  235.70    \\  
8.9 & 7.26e-07 & 2.53e-05 & 34.83   & 0           & 0           & NaN     \\ \hline
\end{tabular}

\vspace{1em}

\subcaption{$n=100$, $p=50$, $k=4$}
\label{my-label}
\begin{tabular}{ccccccccc}
$x$   & $\mathrm{EST_{IS}}$       & $\mathrm{SD_{IS}}$       & $\mathrm{SD_{IS}/EST_{IS}}$ & $\mathrm{EST_{DMC}}$ & $\mathrm{SD_{DMC}}$ & $\mathrm{SD_{DMC}/EST_{DMC}}$    \\ \hline  
10.6 & 7.60e-03    & 5.65e-02     & 7.43   &  8.01e-03 & 8.91e-02 & 11.13 \\  
10.8 & 6.58e-04 & 6.63e-03 &  10.08 & 8.44e-04 &  2.90e-02 &  34.41 \\  
11.0 &  5.49e-05 & 1.47e-03 & 26.73  & 6.40e-05  & 8.00e-03 & 125.00 \\  
11.3 & 1.70e-07 & 5.56e-06 & 32.77  & 0           & 0           & NaN     \\ \hline
\end{tabular}
\end{table}

\begin{table}[!htbp]
\centering
\caption{$U_{n,p}^k$ Results for $\beta=2$.} \label{testrek2}

\subcaption{$n=100 $, $p=50$, $k=2$}
\label{my-label}
\begin{tabular}{ccccccccc}
$x$   & $\mathrm{EST_{IS}}$       & $\mathrm{SD_{IS}}$       & $\mathrm{SD_{IS}/EST_{IS}}$ & $\mathrm{EST_{DMC}}$ & $\mathrm{SD_{DMC}}$ & $\mathrm{SD_{DMC}/EST_{DMC}}$    \\ \hline  
5.6 & 4.67e-03 & 3.52e-02& 7.54 & 5.03e-03 & 7.07e-02  & 14.07  \\  
5.7 & 5.08e-04 & 5.95e-03 & 11.72 & 4.98e-04 & 2.23e-02 &  44.80 \\  
5.8 & 4.75e-05 & 9.55e-04 & 20.12  &3.80e-05 &6.16e-03 & 162.23 \\  
6.0 & 7.71e-08 & 2.48e-06 & 32.18   & 0           & 0           & NaN     \\ \hline
\end{tabular}

\vspace{1em}

\subcaption{$n=100 $, $p=50$, $k=3$}
\label{my-label}
\begin{tabular}{ccccccccc}
$x$   & $\mathrm{EST_{IS}}$       & $\mathrm{SD_{IS}}$       & $\mathrm{SD_{IS}/EST_{IS}}$ & $\mathrm{EST_{DMC}}$ & $\mathrm{SD_{DMC}}$ & $\mathrm{SD_{DMC}/EST_{DMC}}$    \\ \hline  
8.1 &1.78e-03 & 2.08e-02 & 11.67 & 2.16e-03 & 4.64e-02  & 21.50 \\  
8.2 & 3.67e-04 & 8.31e-03 & 22.67 & 2.90e-04 & 1.70e-02 & 58.71 \\  
8.3 & 1.87e-05 & 3.73e-04 & 19.96   & 2.50e-05  & 5.00e-03 &   200.00    \\  
8.5 & 1.50e-07 & 6.90e-06 & 45.95   & 0           & 0           & NaN     \\ \hline
\end{tabular}

\vspace{1em}

\subcaption{$n=100 $, $p=50$, $k=4$}
\label{my-label}
\begin{tabular}{ccccccccc}
$x$   & $\mathrm{EST_{IS}}$       & $\mathrm{SD_{IS}}$       & $\mathrm{SD_{IS}/EST_{IS}}$ & $\mathrm{EST_{DMC}}$ & $\mathrm{SD_{DMC}}$ & $\mathrm{SD_{DMC}/EST_{DMC}}$    \\ \hline  
10.4 & 2.49e-03    & 4.78e-02 & 19.18   & 2.73e-03 & 5.22e-02 & 19.12 \\  
10.5 & 4.27e-04 & 6.15e-03 & 14.40 & 4.42e-04 &  2.10e-02 &  47.55 \\  
10.6 &  5.47e-05 & 1.86e-03 & 34.04  & 6.90e-05  & 8.31e-03 & 120.38 \\  
10.8 & 3.17e-07 & 1.23e-05 & 38.96  & 0           & 0           & NaN     \\ \hline
\end{tabular}
\end{table}

 \section{Conclusions and extensions}\label{sec:ext}
 
 This paper proposes an asymptotically efficient Monte Carlo  method to  estimate the tail probabilities of the ratio of the largest eigenvalue to the trace of the  Wishart matrix. Theoretically, we prove  {that} the importance sampling estimator is  {asymptotically} efficient. Numerically, we conduct extensive studies to evaluate the performance of the proposed algorithm compared  with other existing methods in terms of estimation accuracy  and  computational cost in estimating the tail probabilities. 

The method can be adapted to estimate tail probabilities of the ratio of the sum of the  first $k$ largest eigenvalues  to the trace of the  Wishart matrix, which is defined as
\begin{eqnarray*}
	U_{n,p}^k=\frac{\lambda_1 + \cdots + \lambda_k }{(\lambda_1 + \cdots + \lambda_p)/\min (p,n)},
\end{eqnarray*} where $k$ is a fixed positive integer. 
The algorithm is as follows. First sample $\lambda_2,\ldots,\lambda_p$ from $\LL_{n-1,p-1,\beta}/n$ using the same method  {as} in Algorithm \ref{algo:1}. Second, conditioning on $\lambda_2,\ldots,\lambda_p$, sample $\lambda_1$ from a truncated exponential distribution  {of} the same form as \eqref{eq:expodist}, but redefine 
$$
\tilde{x}=\frac{1}{p-x} \left( x\sum_{i=2}^p \lambda_i-p\sum_{i=2}^k \lambda_i\right)
$$ 
and choose $r$ to be a small constant that  {depends} on the large deviation result of the largest $k$ eigenvalues. 

 We conducted a numerical study to show the  {validity} and efficiency of the proposed method in estimating the tail probabilities of $U_{n,p}^k$. Following the design in Section \ref{sec:num}, the sampling was repeated $10^4$ times for the importance sampling method and $10^6$ times for the direct Monte Carlo method. 
The constant $k$ was chosen to be $2,3,4$, and we took $n=100$, $p=50$, and $r=1/10$. 
Tables \ref{testrek1} and \ref{testrek2}  summarize the results of $\beta=1$ and $\beta=2$, which show similar patterns as Tables \ref{testre1} and~\ref{testre2}. When the tail probability becomes smaller, $SD_{IS}/EST_{IS}$ is smaller than $SD_{DMC}/EST_{DMC}$, which indicates that the importance sampling  is more efficient than the direct Monte Carlo method in estimating the tail probabilities, as discussed in Section \ref{sec:num}. 
It would be interesting to study the asymptotic property of this algorithm on estimating the tail probability of $U_{n,p}^k$. However,  {this would require the} development of  asymptotic theory on the tail probabilities of the first $k$ largest eigenvalues, which is beyond the scope of this  study. We leave it for future work.

\bigskip
 
\section{Proof of Theorem \ref{thm:mainthm}} \label{sec:proof}

This section provides the proof of Theorem \ref{thm:mainthm} on the estimator's asymptotic efficiency. We focus on the case when  $p \leq n$ and $p/n\to \gamma\in (0,1]$. For the case of $p\geq n$ and $p/n\to \gamma\in [1,\infty)$, the proof  follows from the same argument by switching the labels of $n$ and $p$, as shown in Remark \ref{rm:pnswitch}.  

 Recall the definition of $Q$, $ L_{n,p}= \mathbf{1} {(U_{n,p}> x)}dP/dQ$ and $\alpha_{n,p}(x)=\Pr(U_{n,p}>x)$.  
To prove the asymptotic efficiency defined in \eqref{logeff}, we need only show that $\liminf_{n\to\infty}\ln \mathrm{E}_Q ( L_{n,p}^2 )/\{2\ln \alpha_{n,p}(x)\}
\geq  1$ since $\mathrm{E}_Q ( L_{n,p}^2 )/\{2\ln \alpha_{n,p}(x)\}\leq \mathrm{var}_Q ( L_{n,p}^2 )/\{2\ln \alpha_{n,p}(x)\}$. We give an outline of the proof first.
\begin{enumerate}[Step 1.]
	\item We give the asymptotic approximation of $\lim_{n \rightarrow \infty} n^{-1} \ln \alpha_{n,p} (x) = -\gamma I_{\beta}(\beta x)$, where $I_{\beta}(\beta x)$ is the  large deviation rate function. 
	\item By the result in Step 1, we only need to prove that $$\liminf_{n\to\infty}\frac{\ln \mathrm{E}_Q (L_{n,p}^2 )}
{2\ln \alpha_{n,p}(x)}=\liminf_{n\to\infty}\frac{\ln \mathrm{E}_Q (L_{n,p}^2 )}
{-2\gamma I_{\beta}(\beta x)} \geq 1.$$
This is established  using the upper bound $I_1+I_2+I_3$ of $\mathrm{E}_Q(L_{n,p}^2)$ in \eqref{eq:upperboundproof} together with the limiting properties of $I_1,I_2$, and $I_3$ in \eqref{Street_light}, \eqref{Street_light2}, and \eqref{Street_light3}, respectively. 
 \end{enumerate}
 The details of Steps 1 and 2 are given below.

\bigskip
\noindent {\bf Step 1}.  We first obtain the large deviation rate function for $U_{n,p}$, which gives an approximation to $n^{-1}\ln \alpha_{n,p}(x)$ as in \cite{anderson2010introduction}. 
From the argument in \cite{bianchi2011performance}, the large deviation of $U_{n,p}$ has a similar rate function as $\lambda_1$. The explicit form of the large deviation rate function of $\lambda_1$ can be obtained from Theorem 2.6.6 in \cite{anderson2010introduction}. In particular, denote $(\tilde{\lambda}_1,\ldots,\tilde{\lambda}_p)$ to be the unordered eigenvalues of $\mathbf{X}^{H}\mathbf{X}/n$; then from \eqref{originaldensity}, $(\tilde{\lambda}_1,\ldots,\tilde{\lambda}_p)$ has joint density function
\begin{eqnarray*}
f_{n,p,\beta}(\tilde{\lambda}_{1},\ldots,\tilde{\lambda}_{p})  = \frac{1}{p!}  \, C_{n,p,\beta}\prod_{1\leq i<j\leq p}|\tilde{\lambda}_i-\tilde{\lambda}_j|^{\beta}
\times \prod_{i=1}^p \tilde{\lambda}_i^{{\beta(n-p+1)}/{2}-1}\times 
e^{-n \sum_{i=1}^p \tilde{\lambda}_i/2} =  (Z^p_{V,\beta} )^{-1}  | \Delta_{p} (\tilde{\lambda}  )   |^{\beta} e^{-p \sum_{i=1}^p V(\tilde{\lambda}_i)},
\end{eqnarray*}
where the last line follows the notation of (2.6.1) in \cite{anderson2010introduction} with  $\Delta_{p} (\tilde{\lambda}) = \prod_{1\leq i < j \leq p} (\tilde{\lambda}_i-\tilde{\lambda}_j  )$, $ Z_{V,\beta}^{p}  =  p! C_{n,p,\beta}^{-1}$ \ and 
$$V(x) =  \frac{n}{2p}  x   - \frac{\beta( n-p+1)-2}{2p} \ln x ~\sim~ \frac{1}{2} \left \{  \frac{x}{\gamma} -\beta \left( \frac{1}{\gamma} -1 \right)  \ln x \right\}.$$
The notation ``$a_n\sim b_n$" means $a_n = \{ 1+o(1)\} b_n$.
Following the definition in (2.6.3) of \cite{anderson2010introduction}, we further define
\begin{eqnarray*}
	Z_{pV/(p-1),\beta}^{p-1} & = & \int_{\mathbb{R}} \cdots \int_{\mathbb{R}}  | \Delta_{p-1} (\tilde{\lambda}  )   |^{\beta} e^{-(p-1)\sum_{i=1}^{p-1} \left\{ pV(\tilde{\lambda}_i) /(p-1)\right\}} \prod_{i=1}^{p-1} d \tilde{\lambda}_i \\
	&=& \int_{\mathbb{R}} \cdots \int_{\mathbb{R}} \prod_{1\leq i < j \leq (p-1)}  |\tilde{\lambda}_i-\tilde{\lambda}_j   |^{\beta}  \prod_{i=1}^{p-1} \tilde{\lambda}_i^{ {\beta(n-p+1)}/{2}-1} \times e^{-n\sum_{i=1}^{p-1} \tilde{\lambda}_i/2}\prod_{i=1}^{p-1} d \tilde{\lambda}_i.
\end{eqnarray*}
Then, the density function   \eqref{density_l} implies that the normalization constant $Z_{pV/(p-1),\beta}^{p-1}$ equals
$$ 
Z_{pV/(p-1),\beta}^{p-1} = ~\left \{  \frac{1}{(p-1)!} \left(\frac{n}{n-1}\right)^{ {\beta(n-1)(p-1)}/{2}}C_{n-1,p-1,\beta}   \right\} ^{-1}.$$

With the above notation,  Theorem 2.6.6 in \cite{anderson2010introduction} states that the large deviation approximation of $\lambda_{1}=\max( \tilde{\lambda}_1, \ldots, \tilde{\lambda}_p)$ has speed $p$ and good rate function, viz.
\begin{align*}  
I_{\beta}(s) &=
\left\{
\begin{array}{ll}
 -\beta \int_{\mathbb{R}} \ln|s-t| \sigma_{\beta}(dt) +V(s) + \alpha_{V,\beta} & \mbox{ if } s\geq s^*,\\
\infty & \mbox{ if } s<s^*,
\end{array}
\right.     
\end{align*}
 where $s_*=\beta(1-\sqrt{\gamma})^2$,   $s^*=\beta(1+\sqrt{\gamma})^2$, $\sigma_{\beta}(\cdot)$ is the probability distribution function of the  Marchenko--Pastur law specified in \eqref{sigmabeta} and 
$$ 
\alpha_{V,\beta} =- \lim_{p\rightarrow \infty} \frac{1}{p} \ln \frac{Z_{pV/(p-1),\beta }^{p-1}}{Z_{V,\beta}^{p}} .
$$
A direct calculation gives that for $p/n\to\gamma$,
\begin{align}\label{logA3}
\ln \frac{Z_{pV/(p-1),\beta}^{p-1}}{Z_{V,\beta}^p}
&\sim \frac{\beta(n+p)}{2}\ln n-\frac{\beta p}{2}\ln p 
-\frac{\beta n}{2}\ln n-\frac{\beta(p+ n)}{2}\times (\ln \beta-1) +O(\ln n) \nonumber \\
&\sim  \frac{\beta}{2} \left\{ \gamma\ln\left({1}/{\gamma} \right)-(\gamma+1)(\ln \beta -1) \right\} \, n+o(n).
\end{align}
Then, we obtain
$ \alpha_{V,\beta}  =(\beta/2) \times \left\{ \ln\gamma+\left(1/\gamma +1\right)(\ln\beta-1)\right\}.$
Therefore,   the large deviation approximation of $\lambda_{1} = \max( \tilde{\lambda}_1, \ldots, \tilde{\lambda}_p)$ has speed $p$ and rate function
\begin{align} \label{largedevrate}
I_{\beta}(s) 
&=\left\{
\begin{array}{ll}
-\beta  \int_{\mathbb{R}} \ln|s-t| \sigma_{\beta}(dt) + {s}/{(2\gamma)} - ({\beta}/{2}) \left( {1}/{\gamma}-1 \right)\ln s\quad &\\
\quad\quad + ({\beta}/{2}) \left\{ \ln \gamma+\left({1}/{\gamma}+1 \right)(\ln \beta-1) \right\} & \mbox{ if } s\geq s^*, \\
\infty & \mbox{ if }  s<s^*.
\end{array}
\right. 
\end{align}

Recall the notation in Remark \ref{remark:constantdiff} and from result in \cite{bianchi2011performance}, we know when $\mathbf{X}$ has iid  entries $\mathcal{N}(0,1)$ or $\mathcal{CN}(0,1)$, the largest eigenvalue $\bar{\lambda}_1$ and the ratio $U_{n,p}$ defined in (\ref{def:u}) of ${\mathbf{X}^{H} \mathbf{X}}/n$ have the same large deviation approximation function \eqref{largedevrate}. But now in the complex case,  $\mathbf{X}$ has iid  entries $\mathcal{CN}(0,2)$ with $\beta=2$. Similar to the argument in Remark \ref{remark:constantdiff}, since $U_{n,p}$ is invariant to this change,   we have  
$$
\lim_{n\rightarrow \infty} \frac{1}{n} \ln \Pr (U_{n,p}>x) = \lim_{n\rightarrow \infty} \frac{1}{n} \ln \Pr (\bar{\lambda}_{1}>x) 
= \lim_{p\to\infty}\frac{p}{n}\times \frac{1}{p}\ln \Pr (\lambda_{1}> \beta x)= -\gamma I_{\beta}(\beta x).
$$
Therefore we have the large deviation result:
\begin{eqnarray} \label{largerate}
	n^{-1} \ln \alpha_{n,p}(x)\sim - \gamma I_{\beta}\left({\beta x}\right).
\end{eqnarray}
 
\noindent {\textbf{Step 2}}. We focus on $\ln \{ \mathrm{E}_Q( L_{n,p}^2 )\}$ in this step.
 Recall that $\sigma_{\beta}$ in \eqref{sigmabeta} denotes the equilibrium  measure for the large deviations of the empirical distribution of eigenvalues $(\lambda_1,\ldots,\lambda_p)$ under $P$;  see Lemma 2.6.2 from \cite{anderson2010introduction}. Define $t_1$ as a constant such that $t_1 >{n}/(n-1)$ but close to $n/(n-1)$. Let $B(\epsilon)$ be the ball of probability measures defined on $[0,t_1 M]$ with radius $\epsilon$ around
 $\sigma_\beta$ under the following metric $\rho$ that generates the weak convergence of probability measures on $\mathbb{R}$. For two probability
measures  $\mu$ and $\nu$ on $\mathbb{R}$,
\begin{eqnarray}\label{Newton}
\rho(\mu,\nu)=\sup_{\|h\|_L\leq 1}\Big| \int_{\mathbb{R}} h(x)\mu(dx)-\int_{\mathbb{R}} h(x) \nu(dx)
\Big|,
\end{eqnarray}
where $h$ is a bounded Lipschitz function defined on $\mathbb{R}$ with
$$
\|h\| = \sup_{x\in\mathbb{R}} |h(x)|, \quad \|h\|_L=\|h\| +\sup_{x\neq y}|h(x)-h(y)|/|x-y|.
$$
Let ${\cal L}^Q_{p-1}$ be the empirical measure of  $(\lambda_2^*,\ldots,\lambda_p^*)$ with $(\lambda_2,\ldots,\lambda_p)= \{(n-1)/n\} \times (\lambda_2^*,\ldots,\lambda_p^*)$ being constructed as in Step~1 of Algorithm~\ref{algorithm1} under the change of measure $Q$. 

We  know from the Marchenko--Pastur law that ${\cal L}^Q_{p-1} \rightarrow \sigma_{\beta} $ a.s., as defined in (\ref{sigmabeta}). Then
for a large constant  $M$, we have the following upper bound for $\mathrm{E}_Q( L_{n,p}^2)$
 \begin{align}
\mathrm{E}_Q  ( L_{n,p}^2  ) &\leq  \mathrm{E}_Q\left \{  ( {dP}/{d Q} )^2: \lambda_{1}>M\right\} \notag \\
&+\mathrm{E}_Q\left\{ ({dP}/{dQ} )^2: U_{n,p}>x, M>\lambda_1,
 {\cal L}^Q_{p-1}  \notin B(\epsilon)\right\} + \mathrm{E}_Q\left\{ ({dP}/{dQ})^2: U_{n,p}>x, M>\lambda_1,
 {\cal L}^Q_{p-1}  \in B(\epsilon)\right \} \notag \\
 &\equiv  I_1+I_2+I_3. \label{eq:upperboundproof}
\end{align}
We will show that the first two terms of the above upper bound  {are} ignorable, i.e., for any $\epsilon >0$,
  \begin{align}\label{Street_light}
\lim_{M\to\infty}\limsup_{n\to\infty}
\frac{1}{n}\ln I_1 
&=-\infty,\\
\lim_{M\to\infty}\limsup_{n\to\infty} \frac{1}{n}\ln I_2  &= -\infty,
 \label{Street_light2}
\end{align}
and we will further show that
\begin{eqnarray} \label{Street_light3}
 \lim_{\epsilon\to 0, M\to\infty}\limsup_{n\to\infty}\
\frac{1}{n}\ln I_3 =-2\gamma I_{\beta}(\beta x).
\end{eqnarray}
Combining \eqref{Street_light}, \eqref{Street_light2} and \eqref{Street_light3} together, we will then deduce that
\begin{eqnarray*}
	\limsup_{n\to\infty}\
\frac{1}{n}\ln \mathrm{E}_Q(L_{n,p}^2) \leq -2\gamma I_{\beta}(\beta x).
\end{eqnarray*} 
Then by the result in Step~1 of the proof, and the fact that $\ln \alpha_{n,p}(x)<0$, we will conclude that
\begin{eqnarray*}
	\liminf_{n\to\infty}\frac{\ln \mathrm{E}_Q (L_{n,p}^2 )}
{2\ln \alpha_{n,p}(x)}\geq 1.
\end{eqnarray*}

Based on the argument above, in the following we need only prove \eqref{Street_light}--\eqref{Street_light3}.

\bigskip
\noindent
\textit{Proof of \eqref{Street_light}.} 
Let   
$
B_{n,p,\,\beta} =  {Z_{pV/(p-1),\,\beta}^{p-1}}/{Z_{V,\,\beta}^p}.
$
From the construction of the change of measure $Q$, we can  rewrite the left-hand side display in \eqref{Street_light} as
  \begin{align*}
\lim_{M\to\infty}\limsup_{n\to\infty}
\frac{1}{n}\ln 
\mathrm{E}_Q & \biggr [
\biggr\{ \frac{B_{n,p,\,\beta}  \prod_{i=2}^p(\lambda_{1}-\lambda_{i})^{\beta}
\times \lambda_{1}^{{\beta(n-p+1)}/{2}-1}\times 
e^{-n\lambda_{1}/2}}
{nr e^{-nr (\lambda_{1}-\tilde x\vee\lambda_{2})}\times  \mathbf{1}_{(\lambda_{1}>\tilde x\vee \lambda_{2})}}\biggr\}^2:
~ \lambda_{1}>M\biggr]
\notag\\
&\leq
\lim_{M\to\infty}\limsup_{n\to\infty}\frac{1}{n}\ln  
\int_{\lambda_1>M,\atop \lambda_1>\lambda_{2}}
r^{-2}{n^{-2}B^2_{n,p,\beta} \lambda_1^{{\beta(p+n-1)}-2}e^{-n\lambda_1+2rn (\lambda_1-\tilde x\vee\lambda_{2})}}\\
&\quad\quad\quad\quad\quad\quad\quad\quad
\times rne^{-rn (\lambda_1-\tilde x\vee\lambda_{2})} f^Q_{n,p}(\lambda_{2},\ldots,\lambda_{p})
  d\lambda_1  \cdots d \lambda_p  \\
  &\leq
\lim_{M\to\infty}\limsup_{n\to\infty}\frac{1}{n}\ln \int_{\lambda_1>M}{r^{-1}}
n^{-1}B^2_{n,p,\beta}
 \lambda_1^{{\beta(p+n-1)}-2}\times
e^{-n\lambda_1+rn \lambda_1-rn\tilde x}  d\lambda_1.
  \end{align*}
Next we change variable $\lambda_1$ to $\lambda_1+M$ and 
since
$\left(  \lambda_1 +M \right) ^{\beta(p+n-1)-2}$ $\leq M^{\beta(p+n-1)-2} e^{ \left\{ \beta( p+n-1)-2 \right\} \lambda_1/M}$,
  we obtain the following upper bound for the expectation in Eq.~\eqref{Street_light}: 
  \begin{multline*}
\lim_{M\to\infty}\limsup_{n\to\infty}\frac{1}{n}\ln
\int_{0}^\infty
r^{-1}n^{-1}B^2_{n,p,\beta}
M^{{\beta(p+n-1)}-2}
e^{\{ {\beta(p+n-1)}-2\} \lambda_1/M-(n-rn) (\lambda_1+M)-rn \tilde x}
  d\lambda_1\\
=
\lim_{M\to\infty}\limsup_{n\to\infty}\frac{1}{n}\ln
\{ B^2_{n,p,\beta}
M^{{\beta(p+n-1)}-2}
e^{-(n-rn ) M-rn \tilde x} \} +o(1)
 =-\infty,
  \end{multline*}
where  the last step follows from the approximation of $B_{n,p,\,\beta} $ from (\ref{logA3}).
This proves Eq.~\eqref{Street_light}. \hfill $\Box$

\bigskip
\noindent
\textit{Proof of \eqref{Street_light2}.} Consider the expectation term in Eq. \eqref{Street_light2}.
 Since $\lambda_1-\lambda_i< M$ and $\lambda_{2}\vee\tilde x\geq \tilde x $,   the following inequality  holds for any $\epsilon>0$, 
 \begin{eqnarray}
\limsup_{n\to\infty}
\frac{1}{n}\ln I_2
\leq
\limsup_{n\to\infty}\frac{1}{n}\ln \mathrm{E}_Q\biggr[
\biggr\{ \frac{B_{n,p,\,\beta}
 M^{\beta(p-1)}
 \lambda_1^{ {\beta(n-p+1)}/{2}-1}
e^{-n \lambda_1/2}}
{rn e^{-rn (\lambda_1-\tilde x)}}
\biggr \}^2: 
U_{n,p}>x, M>\lambda_1,  {\cal L}^Q_{p-1} \notin B(\epsilon)\biggr].
 \label{aaaa}
\end{eqnarray}
 Under the assumption that $p/n\to\gamma$, $\lambda_1<M$ and with the result from (\ref{logA3}), we know that
 $$
 \frac{B_{n,p,\,\beta}
 M^{\beta(p-1)}
 \lambda_1^{ {\beta(n-p+1)}/{2}-1}
e^{-n \lambda_1/2}}
{rn e^{-rn (\lambda_1-\tilde x)}}=e^{O(nM)}.$$
This implies that
$$
\eqref{aaaa}
\leq \limsup_{n\to\infty}n^{-1} \ln \Big [e^{O(n M)}Q\{ U_{n,p}>x, M>\lambda_1,  {\cal L}^Q_{p-1}  \notin B(\epsilon)\} \Big ]
\leq \limsup_{n\to\infty}\Big [ O(M)+ n^{-1} \ln \Pr \{{\cal L}^Q_{p-1}  \notin B(\epsilon)\}\Big].
$$
The large deviation result for ${\cal L}^Q_{p-1} $ \citep[Theorem 2.6.1 in][]{anderson2010introduction} then yields
$$
\limsup_{n\to\infty}\frac{1}{n^2}  \ln \Pr\{ {\cal L}^Q_{p-1}  \notin B(\epsilon)\} =\limsup_{n\to\infty}\frac{(p-1)^2}{n^2} \times \frac{1}{(p-1)^2} \ln \Pr\{ {\cal L}^Q_{p-1}  \notin B(\epsilon)\}   <0.
$$
This proves \eqref{Street_light2}. \hfill $\Box$

\paragraph{Proof of \eqref{Street_light3}}  Define  $\Omega_n = \{U_{n,p}>x, M>\lambda_1\ \mbox{and} \ {\cal L}^Q_{p-1}  \in B(\epsilon)\}$.
We can write
\begin{eqnarray*}
	I_3 = O(1) n^{-2} B_{n,p,\,\beta}^2 \mathrm{E}_Q \left \{ e^{2\beta \sum_{i=2}^p \ln(\lambda_1-\lambda_i)  }   \lambda_1^{\beta(n-p+1)-2} e^{-n\lambda_1} e^{2nr(\lambda_1- \tilde{x} \vee \lambda_2 )}: \Omega_n  \right \}. 
\end{eqnarray*}
Let $\Phi(z,\epsilon)=\sup_{\mu\in B(\epsilon)}\int \ln(|z-y|)\{ \mu(dy)-\sigma_{\beta}(dy)\}$, we have 
\begin{align*}
\sum_{i=2}^p \ln(\lambda_{1}-\lambda_{i})
& =  (p-1)\int_{\mathbb{R}}\ln\, \biggr(\frac{n\lambda_{1}}{n-1}-y\biggr)\,{\cal L}^Q_{p-1} (dy)-(p-1)\ln \frac{n}{n-1}\\
&\leq  (p-1) \Phi\,\biggr(\frac{n\lambda_{1}}{n-1} ,\epsilon\biggr)
+(p-1)\int \ln\,\biggr(\frac{n\lambda_{1}}{n-1}-y\biggr)\,\sigma_{\beta}(dy)+O(1).
\end{align*}
Under the condition that $\lambda_1<M$,  we know ${n\lambda_1}/(n-1)< 2M$ when $n$ is  {large} enough. Let $G=\max  \{ \beta(1+\sqrt{\gamma})^2,2M  \} $ and define
\begin{eqnarray}\label{df:h}
	h(x)=x \mathbf{1}(x\in \left[0, G\right]).
\end{eqnarray}
Then $h$ is a bounded Lipschitz function on $[0, G]$. 
Furthermore, given $\ {\cal L}^Q_{p-1}  \in B(\epsilon)$ and under measure $Q$, we have
\begin{eqnarray*}\label{eq:concentrate}
\Big|\frac{1}{p-1}\sum_{i=2}^p\frac{n\lambda_i}{n-1} - \beta \Big|  =\Big|  \int_{\mathbb{R}} h(y) {\cal L}^Q_{p-1} (dy) -  \int_{\mathbb{R}} h(y) \sigma_{\beta} (dy) \Big| <  O(\epsilon) ,\end{eqnarray*}
for $\beta \in \{ 1,2\}$. This is because from Theorem 6.3.1 in \cite{dumitriuthesismatrix}, for a distribution with the same density as (\ref{MPlaw2}), the first moment is $\mu_{1,\gamma} = \int \bar{s} \times f(\bar{s})d \bar{s} = 1$. For the density in (\ref{sigmabeta}), similar to  Remark \ref{remark:constantdiff}, the first moment is  $\int s \times \sigma_{\beta}(ds)=\beta \int \bar{s} \times f(\bar{s})d \bar{s} =\beta \times \mu_{1,\gamma}=\beta$. Considering our choice of $G$ in \eqref{df:h}, we have
\begin{eqnarray*} \label{firstmoment}
	\int_{\mathbb{R}} h(y) \sigma_{\beta} (dy) = \int_{\mathbb{R}} y \sigma_{\beta} (dy) = \beta \times \mu_{1,\gamma} = \beta.
\end{eqnarray*}
Therefore, $U_{n,p}>x$ and $\lambda_1>\tilde x$ implies that $\lambda_1>\beta x+O(\epsilon)$ and we can write
$$
I_3
\leq O(1)n^{-1}B_{n,p,\,\beta}^2  
\int_{\beta x+O(\epsilon)}^{M}
e^{2\beta(p-1)\Phi(\frac{n\lambda_1}{n-1},\epsilon)+2\beta(p-1)\int \ln(\frac{n\lambda_1}{n-1}-y)\sigma_{\beta}(dy)}
  \times \lambda_1^{\beta(n-p+1)-2}
e^{-n\lambda_1+rn \{\lambda_1-\beta x+O(\epsilon)\}} d\lambda_1.
$$
Since $\beta x+O(\epsilon) < \lambda_1 < M$, we have $\Phi [n\lambda_{1}/(n-1),\epsilon]
\leq \sup_{z\in[n\{\beta x+O(\epsilon)\}/(n-1), nM/(n-1)]}\Phi(z,\epsilon)$
under the constraint ${\cal L}^Q_{p-1}  \in B(\epsilon)$ and that
\begin{align*}
\int \ln\left(\frac{n\lambda_{1}}{n-1}-y\right)\, \sigma_{\beta}(dy)
&=
\int \ln \left(\frac{n\beta x}{n-1}-y \right )\sigma_{\beta}(dy)
+\int \ln\left\{1+\frac{n\lambda_1-n\beta x}{n\beta x-(n-1)y}\right\} \sigma_{\beta}(dy)\\
&\leq
\int \ln \left( \frac{n\beta x}{n-1}-y \right) \sigma_{\beta}(dy)
+\int  \frac{n\lambda_1-n\beta x}{n\beta x-(n-1)y} \, \sigma_{\beta}(dy).
\end{align*}
It follows that
\begin{multline*}
I_3 \leq O(1)n^{-1} B_{n,p,\,\beta}^2 \times
e^{2 \beta(p-1)  \sup_{z\in [\frac{n(\beta x+O(\epsilon))}{n-1}, \frac{nM}{n-1}]}\Phi(z,\epsilon)
+2\beta(p-1)\int \ln(\frac{n\beta x}{n-1}-y)\sigma_{\beta}(dy)}
\\
  \times \int_{\beta x+O(\epsilon)}^{M}
e^{2 \beta(p-1) \int \frac{n\lambda_1-n\beta x}{n\beta x-(n-1)y}d\sigma_{\beta}(y)}
  \lambda_1^{\beta(n-p+1)-2}
e^{-n\lambda_1+rn \{\lambda_1-\beta x+O(\epsilon)\}} d\lambda_1.
\end{multline*}
The right-hand side equals
\begin{multline*}
 O(1)n^{-1} B_{n,p,\,\beta}^2  \times
e^{2\beta(p-1)  \sup_{z\in[\frac{n(\beta x+O(\epsilon))}{n-1}, \frac{nM}{n-1}]}\Phi(z,\epsilon)
+2\beta(p-1)\int \ln(\frac{n\beta x}{n-1}-y)\sigma_{\beta}(dy)}
\\
\times \int_{O(\epsilon)}^{M-\beta x}
e^{2\beta(p-1) \int \frac{n\lambda_1}{n\beta x-(n-1)y}d\sigma_{\beta}(y)}
\times (\lambda_1+\beta x)^{\,\beta(n-p+1)-2 }\times
e^{-(1-r )n(\lambda_1+ \beta x)-rn\{ \beta x+O(\epsilon)\}} d\lambda_1,
\end{multline*}
where we change the variable $\lambda_1$ to $\lambda_1+ \beta x$ for the integral. Then it follows that
\begin{multline}
I_3 \leq
 O(1)n^{-1} B_{n,p,\,\beta}^2 \times
e^{2 \beta (p-1)  \sup_{z\in[\frac{n(\beta x+O(\epsilon))}{n-1}, \frac{nM}{n-1}]}\Phi(z,\epsilon)
+2\beta(p-1)\int \ln(\frac{n\beta x}{n-1}-y)\sigma_{\beta}(dy)} 
 \times~ (\beta x)^{\beta(n-p+1)-2 }e^{- n\{ \beta x+O(\epsilon)\}}
\\
\times~\int_{O(\epsilon)}^{M-\beta x}
e^{2\beta(p-1)  \int \frac{n\lambda_1}{n\beta x-(n-1)y}d\sigma_{\beta}(y)
+\left\{ \beta(n-p+1)-2  \right\} \frac{\lambda_1}{\beta x}
-(1-r )n\lambda_1} d\lambda_1, \quad \quad
\label{xiaoshu}
\end{multline}
 as we used the fact that  $(\lambda_1+\beta x)^{\beta(n-p+1)-2}\leq (\beta x)^{\beta(n-p+1)-2} e^{\{ \beta(n-p+1)-2\} \lambda_1/(\beta x)} $. 

Under $s^* < \beta x$, we can find a finite number $t_0$ such that $s^*  <t_0 x \leq {n\{\beta x+O(\epsilon)\}}/(n-1)$, for small enough $\epsilon$ and large enough $n$. Recall that $t_1 M \geq nM/(n-1)$. 
Next we show that
\begin{eqnarray}\label{High_tower}
\limsup_{\epsilon\to 0}\sup_{z\in[t_0 x,t_1 M]}\Phi(z,\epsilon)\leq 0.
\end{eqnarray}
 For any $z\in [t_0 x,t_1 M]$ and $\mu\in B(\epsilon)$,
let ${\cal S}_1(z)=\{y\in \mbox{supp}(\sigma_\beta)\cup \mbox{supp}(\mu): |z-y|> \eta\}$ and
${\cal S}_2(z)=\{y\in \mbox{supp}(\sigma_\beta)\cup \mbox{supp}(\mu): |z-y|\leq \eta\}$,
 where $\mbox{supp}(\mu)$ is the support of measure $\mu$ and $\eta$ is a small constant such that $\eta<\min\{t_0 x-s^*,1\}$ with $s^*$ defined   in \eqref{sigmabeta}.
 Note that $\mbox{supp}(\sigma_\beta)\subset {\cal S}_1(z).$
 {Given $z\in [t_0 x,t_1 M]$, set $f_z(y) = \ln(|z-y|)$ for $y\in {\cal S}_1(z)$. 
The Lipschitz norms of the set of functions $\{f_z(\cdot);\, z\in [t_0x,t_1 M]\}$ on ${\cal S}_1(z) $ are bounded by a constant $C<\infty$. By the definition of $\rho(\cdot, \cdot)$ in \eqref{Newton}, we obtain
\begin{align*}
\sup_{z\in [t_0x,t_1 M]}\int_{\mathbb{R}} \ln(|z-y|)\{ \mu(dy)-\sigma_{\beta}(dy)\} 
&\leq 
\sup_{z\in [t_0x,t_1 M]}\int_{{\cal S}_1} f_z(y)\{ \mu(dy)-\sigma_{\beta}(dy)\}
+\sup_{z\in [t_0 x,t_1 M]}\int_{{\cal S}_2} f_z(y)\mu(dy)
\\
&\leq
\sup_{z\in [t_0x,t_1  M]}\int_{{\cal S}_1} f_z(y)\{\mu(dy)-\sigma_{\beta}(dy)\} \leq  C \rho(\mu, \sigma_{\beta})<C\epsilon,
\end{align*}
for any $\mu\in B_{\epsilon}$. This implies that $\sup_{z\in[t_0 x,t_1 M]}\Phi(z,\epsilon)<C\epsilon$. Then \eqref{High_tower} follows.}
When $r<1-2\beta \gamma \int \{1/(\beta x-y)\} d\sigma_{\beta}(y)
-(1-\gamma)/ x$,
we know that the integral term in \eqref{xiaoshu} is $\sim e^{nO(\epsilon)}.$
Therefore
$$
\lim_{\substack{\epsilon\to0 \\ M\to\infty}}\limsup_{n\to\infty}\frac{1}{n}\ln I_3 
=  2 \beta \gamma \int  \ln(\beta x-y)\sigma_{\beta}(dy)-\beta x+\beta(1- \gamma)\ln(\beta x)  - \beta \left\{  \gamma\ln \gamma +(1+\gamma)\left(\ln \beta-1 \right) \right\}  = - 2\gamma I_{\beta}(\beta x),
$$
where  $I_{\beta}(x)$ is defined as in \eqref{largedevrate}. Therefore, we conclude that
\begin{eqnarray*}
&&\limsup_{n \to\infty} \ln
 \mathrm{E}_Q (L_{n,p}^2 )/n \leq -2 \gamma I_{\beta}(\beta x).
\end{eqnarray*}
Hence, the above upper bound and the approximation in \eqref{largerate} imply that
$$
\liminf_{n\to\infty} {\ln \mathrm{E}_Q (L_{n,p}^2 )}/
\{2\ln \alpha_{n,p}(x)\}\geq 1,$$
where note that $\ln \alpha_{n,p}(x)<0 $. 
This completes the proof. \hfill $\Box$

\section*{Acknowledgments} 
The authors thank Prof. Boaz Nadler for suggesting this problem, as well as
 the Editor-in-Chief, an Associate Editor, and two reviewers  for many helpful and constructive comments.
This work was supported in part by 
National Security Agency grant H98230--17--1--0308 and National Science Foundation
 grant DMS--1712717.


\section*{References}
\begin{thebibliography}{10}

\bibitem{anderson2010introduction}
G.W. Anderson, A.~Guionnet, O.~Zeitouni,
{An Introduction to Random Matrices},
Cambridge University Press, 2010.

\bibitem{ASMGLY07}
S.~Asmussen, P.~Glynn,
{Stochastic Simulation: Algorithms and Analysis},
Springer, New York, 2007.

\bibitem{AsmKro06}
S.~Asmussen, D.~Kroese,
Improved algorithms for rare event simulation with heavy tails,
Adv. Appl. Probab. 38 (2006) 545--558.

\bibitem{bianchi2011performance}
P.~Bianchi, M.~Debbah, M.~Ma{\"\i}da, J.~Najim,
Performance of statistical tests for single-source detection using
  random matrix theory,
IEEE Trans. Inform. Theory 57 (2011) 2400--2419.

\bibitem{BlaGly07}
J.~Blanchet, P.~Glynn,
Efficient rare event simulation for the maximum of heavy-tailed random
  walks,
Ann. Appl. Probab. 18 (2008) 1351--1378.

\bibitem{chiani2014distribution}
M.~Chiani,
Distribution of the largest eigenvalue for real Wishart and
  Gaussian random matrices and a simple approximation for the
  Tracy--Widom distribution,
J. Multivariate Anal. 129 (2014) 69--81.

\bibitem{davis1972ratios}
A.~Davis,
On the ratios of the individual latent roots to the trace of a Wishart matrix,
J. Multivariate Anal. 2 (1972) 440--443.

\bibitem{Demmel:1997:ANL:264989}
J.W. Demmel,
Applied Numerical Linear Algebra,
Society for Industrial and Applied Mathematics, Philadelphia, PA, 1997.

\bibitem{dumitriu2002matrix}
I.~Dumitriu, A.~Edelman,
Matrix models for beta ensembles,
J. Math. Phys. 43 (2002) 5830--5847.

\bibitem{dumitriuthesismatrix}
I.~Dumitriu,  A.~Edelman,
Eigenvalue Statistics for Beta Ensembles,
Doctoral dissertation, Massachusetts Institute of Technology, Boston, MA, 2003.

\bibitem{DupLedWang07}
P.~Dupuis, K.~Leder, H.~Wang,
Importance sampling for sums of random variables with regularly
  varying tails,
ACM Trans. Model. Comput. Simul. 17 (2007) 1--14.

\bibitem{xu2016rare}
T.~Jiang, K.~Leder,  G.~Xu,
Rare-event analysis for extremal eigenvalues of white {Wishart}
  matrices, Ann. Statist. 45 (2017) 1609--1637.

\bibitem{johansson2000shape}
K.~Johansson,
Shape fluctuations and random matrices,
Comm. Math. Phys. 209 (2000) 437--476.

\bibitem{johnstone2001}
I.M. Johnstone,
On the distribution of the largest eigenvalue in principal components
  analysis,
Ann. Statist. 29 (2001) 295--327.

\bibitem{karouicomplexrate}
N.~Karouim,
A rate of convergence result for the largest eigenvalue of complex
  white Wishart matrices,
Ann. Probab. 34 (2006) 2077--2117.

\bibitem{kortun2012distribution}
A.~Kortun, M.~Sellathurai, T.~Ratnarajah, C.~Zhong,
Distribution of the ratio of the largest eigenvalue to the trace of
  complex {W}ishart matrices,
IEEE Trans. Signal Proc. 60 (2012) 5527--5532.

\bibitem{krzanowski2000principles}
W.~Krzanowski,
{Principles of Multivariate Analysis},
Oxford University Press, 2000.

\bibitem{kuriki2001tail}
S.~Kuriki, A.~Takemura,
Tail probabilities of the maxima of multilinear forms and their
  applications,
Ann. Statist.  {29} (2001) 328--371.

\bibitem{LiuXuTomacs}
J.~Liu, G.~Xu,
Efficient simulations for the exponential integrals of {H}\"{o}lder
  continuous {G}aussian random fields,
ACM Trans. Model. Comput. Simul.
  24 (2014) 1--9.

\bibitem{LiuXu12}
J.~Liu, G.~Xu,
On the conditional distributions and the efficient simulations of
  exponential integrals of {Gaussian} random fields,
Ann. Appl. Probab. 24 (2014) 1691--1738.

\bibitem{ma2012accuracy}
Z.~Ma,
Accuracy of the Tracy--Widom limits for the extreme eigenvalues
  in white Wishart matrices, Bernoulli 18 (2012) 322--359.

\bibitem{muirhead2009aspects}
R.J. Muirhead, Aspects of Multivariate Statistical Theory,
Wiley, New York, 2009.

\bibitem{nadler2011distribution}
B.~Nadler,
On the distribution of the ratio of the largest eigenvalue to the
  trace of a Wishart matrix,
J. Multivariate Anal. 102 (2011) 363--371.

\bibitem{paul2014random}
D.~Paul, A.~Aue,
Random matrix theory in statistics: A review,
J. Stat. Plann. Inf. 150 (2014) 1--29.

\bibitem{schuurmann1973distributions}
F.~Schuurmann, P.~Krishnaiah, A.~Chattopadhyay,
On the distributions of the ratios of the extreme roots to the trace
  of the Wishart matrix,
J. Multivariate Anal. 3 (1973) 445--453.

\bibitem{SIE76}
D.~Siegmund,
Importance sampling in the Monte Carlo study of sequential tests,
Ann. Statist. 4 (1976) 673--684.

\bibitem{tse2005fundamentals}
D.~Tse, P.~Viswanath,
Fundamentals of Wireless Communication,
Cambridge University Press, 2005.

\bibitem{wei2012exact}
L.~Wei, O.~Tirkkonen, P.~Dharmawansa, M.~McKay,
On the exact distribution of the scaled largest eigenvalue,
In: 2012 IEEE International Conference on Communications (ICC),
  pp. 2422--2426, 2012.

\bibitem{xu2014rare}
G.~Xu, G.~Lin, J.~Liu,
Rare-event simulation for the stochastic {K}orteweg--de {V}ries
  equation, SIAM/ASA J. Uncertainty Quantification 2 (2014) 698--716.

\end{thebibliography}
\end{document}